\documentclass[11pt, a4paper]{article}

\usepackage{xcolor}
\usepackage{float}
\usepackage{verbatim}
\usepackage{fancyvrb}

\let\proglang=\textsf
\newcommand{\pkg}[1]{{\fontseries{b}\selectfont #1}}

\newenvironment{CodeChunk}{}{}
\DefineVerbatimEnvironment{CodeInput}{Verbatim}{fontshape=sl}
\newcommand{\tcb}[1]{\textcolor{black}{#1}}

\usepackage{url}
\usepackage{a4wide}
\usepackage{graphicx}
\usepackage{natbib}
\usepackage[shortlabels]{enumitem}
\usepackage{hyperref}
\usepackage{setspace}
\usepackage{multirow, booktabs}
\usepackage{amsmath, amsthm, amssymb, amsfonts}
\usepackage{bm, bbm}

\usepackage{longtable}
\usepackage{color}
\usepackage{mathtools}
\usepackage{appendix}

\usepackage{authblk}
\usepackage{caption}
\usepackage{subcaption}

\title{TrendLSW: Trend and Spectral Estimation of Nonstationary Time Series in R}

\author[1]{Euan T. McGonigle}
\author[2]{Rebecca Killick}
\author[3]{Matthew A. Nunes}
\affil[1]{ School of Mathematical Sciences, University of Southampton} 
\affil[2]{Department of Mathematics and Statistics, Lancaster University} 
\affil[3]{Department of Mathematical Sciences, University of Bath} 

\begin{document}

\maketitle

\begin{abstract}
The TrendLSW R package has been developed to provide users with a suite of wavelet-based techniques to analyse the statistical properties of nonstationary time series. The key components of the package are (a) two approaches for the estimation of the evolutionary wavelet spectrum in the presence of trend; and (b) wavelet-based trend estimation in the presence of locally stationary wavelet errors via both linear and nonlinear wavelet thresholding; and (c) the calculation of associated pointwise confidence intervals. Lastly, the package directly implements boundary handling methods that enable the methods to be performed on data of arbitrary length, not just dyadic length as is common for wavelet-based methods, ensuring no pre-processing of data is necessary. The key functionality of the package is demonstrated through two data examples, arising from biology and activity monitoring.
\end{abstract}


\section[Introduction]{Introduction}\label{sec:introduction}
Modern time series data can often possess complex characteristics. Given technological advancements in data recording tools, leading to time series being observed over increasingly larger time-scales, it is common for the statistical properties of a time series to vary over time. Examples of so-called `nonstationary' time series can be found in a wide variety of applications areas, including climatology \citep{beaulieu2018distinguishing}, economics \citep{roueff2019time}, and epidemiology \citep{jiang2023time}.

Two key quantities of interest, the mean function and dependence structure, are amongst the most commonly studied properties in time series analysis. Modelling how these properties evolve is crucial for making informed inference on the data: incorrectly assuming stationary behaviour may lead to drawing misleading conclusions. It is known to be a challenging problem to estimate one of these time-varying properties when the other is stationary, and yet more challenging when both display time-dependent characteristics. Therefore, it is common practice for analysis to be restricted to either the mean or dependence.

For mean function estimation in time series, there is a vast literature dedicated to topics including tackling nonparametric regression \citep{von2000non, vogt2012nonparametric}, and trend estimation and detection \citep{wu2007inference, zhang2011testing}. 

For modelling nonstationary dependence structures, a number of approaches have been proposed; for an overview, see \cite{dahlhaus2012locally}. For example, \cite{priestley1965evolutionary} introduces evolutionary processes, \cite{dahlhaus1997fitting} defines the locally stationary Fourier (LSF) processes, whilst \cite{nason2000wavelet} introduce the locally stationary wavelet (LSW) model. \cite{ombao2002slex} consider the smoothed localised exponential (SLEX) model, whilst \cite{van2006semiparametric} discuss nonstationary
autoregressive models, where the parameters are allowed to vary with time. \cite{zhou2009local} define locally stationary time series as an ``input-output" physical system. %as defined in \cite{wu2005nonlinear} with a time-varying input function. 
\tcb{Other approaches which can capture both time-varying means and time-varying dynamics are time-varying structural autoregression models (e.g. \citep{kilian2017structural}) and dynamic linear models, see for example, \cite{keele2006dynamic}.}

Existing software in the area of nonstationary time series predominantly focuses on estimation of either the mean or dependence structure. A comprehensive review of all such packages is outside the scope of this article (in particular any \proglang{R} package implementing a regression approach can be adapted to trend estimation in time series data by using the time index as the covariate).  \tcb{For example, functionality for trend estimation via penalised regression approaches or regression splines (such as in the \pkg{genlasso} \citep{arnold22:genlasso} and \pkg{earth} \citep{milborrow24:earth} packages respectively) could be adapted to serially dependent noise.} For estimation of nonstationary dependence structure, the \proglang{R} package \pkg{LSTS} \citep{olea2015lsts} implements the LSF approach. \cite{nason2010wavethresh} implements wavelet-based methods for time series in the \pkg{wavethresh} package, including the LSW model, as well as wavelet-based trend estimation but for a second-order stationary time series. The \pkg{locits} package \citep{locits} allows for time-varying autocovariance estimation under an LSW model, whilst the \pkg{LSWPlib} library \citep{cardinali22:LSWPlib} provides locally stationary wavelet packet representations and associated spectral estimation methods. \cite{taylor2019multivariate} extend the LSW approach to multivariate time series in the package \pkg{mvLSW}.  Other application-focused packages for nonstationary time series analysis include the \pkg{RSEIS} package \citep{lees23:RSEIS}, which has functionality for spectrogram computation and visualisation for seismic data. For trend detection, packages include \pkg{trend} \citep{pohlert2020trend} and \pkg{funtimes} \citep{lyubchich2023funtimes}. We mention that, for nonparametric regression without a focus on the time series setting, packages include \pkg{mgcv} \citep{wood2023mgcv} and \pkg{gam} \citep{hastie2023gam}. 
\tcb{However, to the best of our knowledge, there are few \proglang{R} packages available specifically designed for estimation of trend in the presence of nonstationary dependence.  Examples include software for dynamic linear models in the \pkg{dynlm} package \citep{zeileis19:dynlm}, also available in the nonlinear time series setting in \pkg{tsDyn} \citep{fabio09:tsDyn, stigler19:nonlinear}.  However, these are not designed for the locally stationary setting.}

In this article we present an \proglang{R} implementation of the Trend Locally Stationary Wavelet (TLSW) model, proposed in \cite{mcgonigle2022trend, mcgonigle2022modelling}. The approach enables the practitioner to estimate both the time-varying mean and dependence structure of a univariate time series
\begin{equation}\label{eq:model1}
X_{t} = T_t + \varepsilon_{t},    \quad 0 \leq t \leq n - 1,
\end{equation}
where $T_t$ represents a smooth deterministic trend function and $\varepsilon_t$ is a mean-zero nonstationary noise term assumed to follow an LSW model, details of which will be specified fully in Section~\ref{sec:trend-lsw}. 

 The \pkg{TrendLSW} package implements the work of \cite{mcgonigle2022modelling,mcgonigle2022trend}, building upon the tools provided in \pkg{wavethresh} to include a nonstationary trend component. This enables the estimation of both first- and second-order properties of a nonstationary time series within the same software package. The \pkg{TrendLSW} package \citep{mcgonigle2024trend} is available from the Comprehensive \proglang{R} Archive Network (CRAN) at \url{https://CRAN.R-project.org/package=TrendLSW}.

 The remainder of the article is organised as follows. A brief background to wavelets and a description of the TLSW model is given in Section~\ref{sec:trend-lsw}. Section~\ref{sec:ews-estimation} describes the estimation of the spectrum and its implementation in \proglang{R}, while Section~\ref{sec:trend-estimation} describes the estimation of the trend and implementation in \proglang{R}. Section~\ref{sec:worked-examples} discusses an extended worked example, and section~\ref{sec:data-example} describes two real data examples that highlight the main functionality of the \pkg{TrendLSW} package. Concluding remarks and discussion are given in Section~\ref{sec:conclusion}.

\section{The Trend-LSW model}\label{sec:trend-lsw}

This section describes the TLSW model of \cite{mcgonigle2022modelling, mcgonigle2022trend} for analyzing nonstationary time series with trend components.

\subsection{Wavelets}

\tcb{Broadly speaking, wavelets are localised, oscillatory basis functions that possess
several useful properties not typically enjoyed by Fourier trigonometric basis functions.} In the LSW framework (discrete) wavelets act as building blocks in an analogous fashion to Fourier exponentials in the classical Cram\'{e}r representation for stationary processes \tcb{and spline bases for mean representations}. Let $\psi$ be a compactly supported wavelet, for example any within the Daubechies family \citep{daubechies1992ten}. Denote by $\{h_k, g_k\}$ the low- /high- pass filter pair associated with $\psi$. The pair  $\{h_k, g_k\}$ filter a signal into low and high frequency components respectively. Letting $N_h$ be the number of non-zero values of $\{h_k \}$, the filters are related by the equation $g_k = (-1)^k h_{N_h-k}$ \citep[Equation (2.52)]{nason2010wavelet}.

Setting $L_j = (2^{j} - 1)(N_h-1)+1$, the discrete wavelets at a given scale $j \in \mathbb{Z}^{+}$, as discussed in \cite{nason2000wavelet}, are defined as the vectors $\psi_j = (\psi_{j,0} , \ldots , \psi_{j, L_j-1})$ of length $L_j$, where 
\begin{align}
\psi_{1,l} &= \sum_k g_{l-2k} \delta_{0k}=g_l,  \quad \quad  l = 0 , \ldots , L_{1} -1 \label{eq:nondec1},\\
\psi_{j+1,l} &= \sum_k h_{l-2k} \psi_{j,k} , \qquad \qquad l = 0 , \ldots , L_j -1 ,\label{eq:nondec2}
\end{align}
where $\delta_{0k}$ is the Kronecker delta. The discrete father wavelet is defined similarly using the associated low-pass filter $\{ h_k \}$. The simplest example of a wavelet basis is the Haar wavelet, which is given by
\begin{equation*}
\psi^{H}_{j,k} = 2^{-j/2} \mathbb{I} \left(0 \leq k \leq 2^{j-1} - 1 \right) - 2^{-j/2} \mathbb{I} \left(2^{j-1}  \leq k \leq 2^{j} - 1 \right),
\end{equation*}
where $j = \{1, 2, 3, \ldots \}$ and $k \in \mathbb{Z}$. 

For a good introduction of the use of wavelets in statistics, we refer the reader to \cite{antoniadis2001wavelet} and \cite{nason2010wavelet}.  \tcb{Other excellent texts on wavelets include \cite{percival2006wavelet} and \cite{vidakovic2009statistical}}.

\subsection{The TLSW process}\label{sec:TLSWprocess}

In this section we define the TLSW process and the key quantities of interest for analysis. A time series $\{X_{t,n} \}_{t=0}^{n-1}$ with $n = 2^J \geq 1$ for $J \in \mathbb{N}$ is said to be a trend locally stationary wavelet (TLSW) process if it admits the representation 
\begin{equation}\label{lsw_rep}
X_{t} = T_t + \varepsilon_t = T_t + \sum_{j = 1}^{\infty} \sum_{k \in \mathbb{Z}} w_{j,k;n} \psi_{j,k} (t) \xi_{j,k} ,
\end{equation}
where $\{\xi_{j,k} \}$ is a random, uncorrelated, zero-mean orthonormal increment sequence, $\{w_{j,k;n} \}$ is a set of amplitudes (parameters to estimate), and $\{\psi_{j, k} \}_{j,k}$ is a set of discrete non-decimated wavelets defined using the discrete wavelets given in Equations~\ref{eq:nondec1} and~\ref{eq:nondec2}. 
That is, there is a time-varying trend component and the error component is a locally stationary wavelet (LSW) process.  The trend component $T_t := T(t/n)$ in Equation~\ref{lsw_rep} is assumed to be a H\"{o}lder continuous function. 
%with constant $K>0$, i.e
% \begin{equation*}
% \left| T \left( \frac{t}{n} \right) -  T \left( \frac{s}{n} \right) \right| \leq \frac{K}{n}, \ \forall \ s,t \in [0,n].
%\end{equation*}
Furthermore, for each $j \geq 1$, there exists a Lipschitz continuous function $W_j(z)$ for $z \in (0,1)$ which satisfies the following properties:
\begin{enumerate}
\item $\sum_{j = 1}^{\infty} | W_j (z) |^2 < \infty$ uniformly in $z \in (0,1)$;
\item the Lipschitz constants $L_j$ are uniformly bounded in $j$ and $\sum_{j=1}^{\infty} 2^{j} L_j < \infty$;
\item there exists a sequence of constants $C_j$ such that for each $n$
\begin{equation*}
\sup_k \left| w_{j,k;n} - W_j \left( \frac{k}{n} \right) \right| \leq \frac{C_j}{n},
\end{equation*}
where for each $j \geq 1$ the supremum is over $k = 0, \ldots , n-1$ and the sequence $\{C_j\}$ satisfies $\sum_{j=1}^{\infty} C_j < \infty$. 
\end{enumerate}

The model imposes the standard assumptions on the LSW component in the literature, allowing for locally stationary dependence structure. Model~\ref{lsw_rep} also permits nonstationary first-order behaviour by incorporating a smooth mean function $T$. Whilst the formal definition requires $n=2^J$, in practice the boundary handling method described in \cite{mcgonigle2022trend} can be used to analyse time series of arbitrary length. Note that there is explicit dependence on $n$ in the model, but this notation is suppressed for clarity of presentation.

As with classical time series theory, the key quantity of interest for characterising dependence structure is the spectrum. The evolutionary wavelet spectrum (EWS) of an LSW process is defined as $S_{j} (z) =  W_{j} (z) ^{2}$ for rescaled time $z = k/n \in (0,1)$ and measures the contribution to variance at a particular rescaled time $z$ and scale $j$. The local autocovariance (LACV) function for a LSW process provides information about the covariance at a rescaled location $z = k/n \in (0,1)$. The LACV, $c(z, \tau)$, of a LSW process with EWS $\{S_j (z) \}$ is defined as $c(z, \tau) = \sum_{j = 1}^{\infty} S_j (z) \Psi_j (\tau)$, for $\tau \in \mathbb{Z}$, $z \in (0,1)$. The LACV is a decomposition of the autocovariance of a process over scales and rescaled time locations, and converges to the process autocovariance $c_n(z, \tau) = \mathbb{E}(\varepsilon_{\lfloor zn \rfloor} \varepsilon_{\lfloor zn \rfloor + \tau} )$ with increasing sample size $n$.

\subsubsection*{Simulating realisations of a TLSW process}

We can simulate an example realisation of a TLSW process with a specified trend and given EWS using \texttt{TLSWsim()}.

The input arguments for \texttt{TLSWsim()} are:
\begin{itemize}
\item \texttt{trend}: the trend component of the time series, which can be either a numeric vector of length $n$, or a function of a single argument on rescaled time $(0,1)$.
\item \texttt{spec}: the spectrum that defines the dependence of the time series. This can be either a numeric $J \times n$ matrix, or a list of length $J$ with each element given by a function of a single argument on rescaled time $(0,1)$ (scales where $S_j (z)=0$ for all $z$ can be represented by the \texttt{NULL} value). For readers familiar with the \texttt{wavethresh} package, this argument can also be given as an object of class \texttt{wd}.
\item \texttt{filter.number}: This specifies the number of vanishing moments of the wavelet in the TLSW model~\ref{lsw_rep} i.e., selects the smoothness of wavelet that you want to use in the decomposition. By default this is $4$.
\item \texttt{family}: specifies the family of wavelets that you want to use in the TLSW model. The options are \texttt{"DaubExPhase"} (default) and \texttt{"DaubLeAsymm"}.
\item \texttt{innov.func}: the function used to generate the innovation sequence $\{\xi_{j,k}\}$.  By default, this is \texttt{rnorm()}.
\end{itemize}

We note that it is also possible to simulate TLSW processes of non-dyadic length $n$, provided that the \texttt{trend} and \texttt{spec} arguments are provided as a numeric vector of length $n$ and numeric matrix of dimensions $\lfloor \log_2 (n) \rfloor \times n$ respectively. The recommended choices for the \texttt{family} argument  are \texttt{"DaubExPhase"} or \texttt{"DaubLeAsymm"}, corresponding to the Daubechies Extremal Phase and Daubechies Least Asymmetric wavelet families respectively. For these families, the \texttt{filter.number} argument can be an integer between $1-10$ and $4-10$ respectively, representing the number of vanishing moments of the wavelet (the Haar wavelet is the Extremal Phase wavelet with 1 vanishing moment).

\begin{figure}[hbt]
\centering
\includegraphics[width =0.9\textwidth]{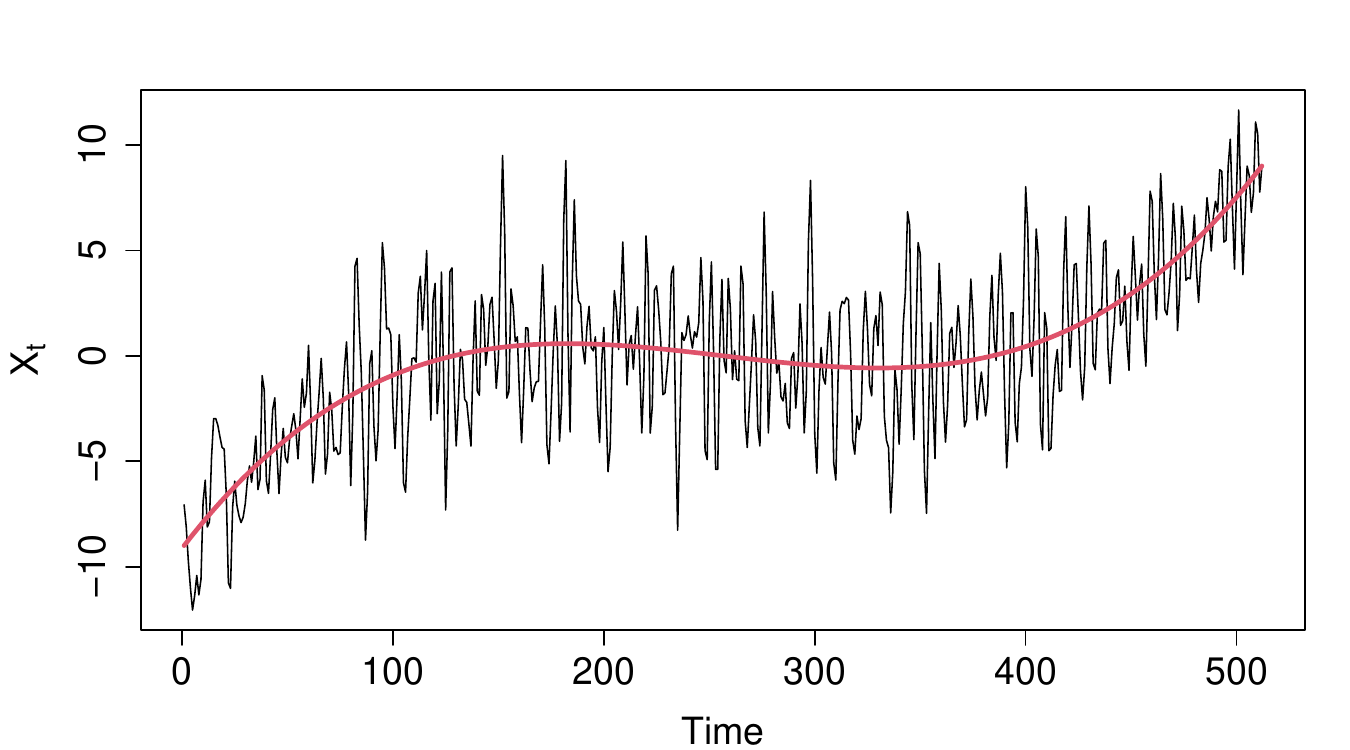}
\caption{Example TLSW process generated with cubic trend function and spectrum with power at scales 2 and 4. Solid red line shows the true underlying trend function.}
\label{fig:example1}
\end{figure}

The example below generates a time series of length $n=512$ where we impose a cubic trend $T_t = 3 (32 (t/n)^3-48 (t/n)^2+22 (t/n)-3)$ and an EWS with a quadratically increasing and decreasing power at level $j=2$, as well as constant power at level $j=4$ i.e., $S_2(t/n) = 2 +12(t/n) -12(t/n)^2$ and $S_4(t/n) = 2$. The specification of the trend is given as a function, and the EWS is defined using a list of functions:

\begin{CodeChunk}
\begin{CodeInput}
R> n1 <- 512
R> trend1 <- function(z){3 * (32 * z^3 - 48 * z^2 + 22 * z - 3)}
R> spec1 <- vector(mode = "list", length = log2(n1))
R> spec1[[2]] <- function(z){2 + 12 * z - 12 * z^2}
R> spec1[[4]] <- function(z){2}
R> set.seed(123)
R> x1 <- TLSWsim(trend = trend1, spec = spec1, filter.number = 4)
\end{CodeInput}
\end{CodeChunk}
Figure~\ref{fig:example1} shows the resulting simulated time series from the TLSW model, where time-varying behaviour in the trend and variability can be seen. 

\subsubsection*{Estimation}\label{sec:ews}

Given an observed time series which we model using (\ref{lsw_rep}), we are interested in estimating the following three quantities: 
\begin{enumerate}
    \item The spectrum (EWS) $\{ S_j (t/n) \}_{j=1}^{J_0}$ for all time locations $t = 0, \ldots, n-1$, with maximum scale of interest $J_0 \leq J$,
    \item The autocovariance (LACV) $c(t/n,\tau)$ for some $\tau$ of interest and $t = 0, \ldots, n-1$ (and associated local autocorrelation $c(t/n,\tau)/c(t/n,0)$),
    \item The trend $T_t$ for $t = 0, \ldots, n-1$.
\end{enumerate}

We consider the estimation of these quantities in the following sections.

The main function within the package is \texttt{TLSW()}, which allows the user to simultaneously estimate both the trend and spectrum of a TLSW process. The package provides two separate, but related, approaches to estimate the trend component of the process. Similarly, there are two approaches implemented for estimating the EWS, each aimed to be used with the corresponding trend estimation.

In Section~\ref{sec:ews-estimation}, we describe the two ways in which EWS estimation can be performed, and in Section~\ref{sec:trend-estimation} we describe the methods for trend estimation. In Section~\ref{sec:worked-examples} we provide worked examples highlighting how these approaches can be applied together using \texttt{TLSW()}.

\section{EWS estimation}\label{sec:ews-estimation}

In what follows, we discuss two approaches for estimating the EWS in the presence of trend. Analogously to classical Fourier methods, we base our estimation procedure on the wavelet periodogram, with some modifications to account for the presence of the trend function. Depending on the properties of the trend function, we give two related but different strategies for estimating the EWS. The general approach for both estimation methods consists of three main steps:

\begin{enumerate}
    \item Calculating the wavelet periodogram using a nondecimated wavelet transform.
    \item Smoothing the wavelet periodogram.
    \item Obtaining the spectral estimate by bias correction of the wavelet periodogram.
\end{enumerate}

Below, for each of the two methods, we describe these three steps in more concrete detail.  The key distinction between the two approaches is that the first directly estimates the EWS accounting for the trend within the estimation. The second approach differences the data and estimates the EWS within the differenced data, then corrects for the effect of the differencing. We first describe the direct estimation approach before detailing the differencing approach.

\subsection{Direct estimation}\label{sec:tlsw-spec1}

To construct the wavelet periodogram, we calculate the nondecimated wavelet coefficients of the time series $\{ X_t\}_{t=0}^{n-1}$. For a given location $k$ and scale $j$, the wavelet coefficient $d_{j,k}$ is defined as $d_{j,k} =  \sum_{t} X_t \psi_{j,k} (t)$. The periodogram is then constructed as $I_{j,k} =  d_{j,k} ^2$. 

If the trend $T(t/n)$ is smooth and does not display irregularities such as cusps or discontinuities, then the wavelet transform coefficients $d_{j,k}$ will be largely free from the effects of the trend. This is due to fact that wavelets naturally act as differencing operators over different time-scales. As mentioned in Section~\ref{sec:TLSWprocess}, wavelets within both the Daubechies Extremal Phase and Least Asymmetric families are constructed to have a given number of vanishing moments. For a wavelet with $m$ vanishing moments, all wavelet coefficients of a polynomial trend function with degree at most $m-1$ will be zero. Therefore, provided the trend function is well-approximated by a polynomial, this useful property of wavelets ensures that the wavelet coefficients, and therefore the wavelet periodogram, are not contaminated by the trend function.

Let the operator $A = (A_{jl})_{j, l >0}$ be given by $A_{jl} = \sum_\tau \Psi_j (\tau) \Psi_l (\tau)$, where the autocorrelation wavelets $\Psi_j ( \cdot )$ are defined by $\Psi_j (\tau) := \sum_{k \in \mathbb{Z}} \psi_{j,k} (0) \psi_{j,k} (\tau)$, for $j > 0, \tau \in \mathbb{Z}$. Assuming we use a wavelet with $m$ vanishing moments and the trend function is polynomial of degree at most $m-1$, then, mirroring a result from \cite{nason2000wavelet}, \cite{mcgonigle2022trend} show that the expectation and variance of the wavelet periodogram are given by
\begin{eqnarray}\label{usual-exp-tlsw}
\mathbb{E} \left( I_{j,k} \right) & \approx& \sum_{l} A_{jl} S_l (k/n),\\
\label{usual-var-tlsw}
\text{Var}( I_{j,k} ) &\approx& 2 \left( \sum_l A_{jl} S_l (k/n) \right)^2 ,
\end{eqnarray}
where `$\approx$' denotes asymptotic equivalence with increasing sample size $n$. This expression also holds for H\"{o}lder continuous trend functions but with a larger finite sample bias than polynomial trends. Equations~\ref{usual-exp-tlsw} and~\ref{usual-var-tlsw} show that the wavelet periodogram is a biased, inconsistent estimator of the EWS. Therefore, the periodogram is first smoothed, then bias corrected, in order to yield a consistent estimator, which motivates steps 2 and 3 above.

The standard approach to smoothing (step 2) is to use some form of linear kernel smoothing.  Here we explicitly describe smoothing using a running mean, but note that the \pkg{TrendLSW} package also implements median smoothing and Epanechnikov kernel smoothing. For bin width size $2N+1$, the smoothed wavelet periodogram is given by
\begin{equation}\label{eq:runmean}
 \widehat{I}_{j,k} = \frac{1}{2N+1} \sum_{m=-N}^{N} I_{j,k+m}.
\end{equation}

Step 3 then bias corrects the smoothed periodogram.  The resulting estimator is given by
\begin{equation}\label{eq:spec-est1}
\widehat{S}_{j} (k/n) = \sum_{l=1}^{J}  A_{lj}^{-1} \widehat{I}_{l,k},
\end{equation}
and the LACV is estimated using $\widehat{S}_{j} (k/n)$ with the equation $\hat{c} (k/n, \tau) = \sum_{j=-J}^{-1} \widehat{S}_{j} (k/n) \Psi_{j} (\tau)$.

The standard approach works well enough in zero-mean cases.  However, the caveat of the above discussion is the occurrence of well-known boundary effects; near the end-points of the time series, where the wavelet filter is of large enough size, wavelet coefficients are computed using reflections of the data, resulting in artefacts. The coarser the scale $j$, the larger the wavelet filter, and therefore the more pronounced these boundary effects are. These are more problematic in datasets containing non-zero trends.  In a practical setting, if the trend is not exactly polynomial, which is often the case in reality, the wavelet coefficients of the trend will not be exactly zero and will be increasingly contaminated at larger scales due to the larger filter length. To mitigate this problem we propose two simple solutions.

The first step involves changing the way that the wavelet transform is performed to improve performance at the boundaries (step 1). Briefly, we use a modification of reflective boundary handling that projects the trend smoothly at the boundaries. This helps to minimise boundary problems and can also be applied to time series of arbitrary length, removing the restrictive assumption in the \pkg{wavethresh} package that the time series length must be dyadic. For full details, see Section 4.4 of \cite{mcgonigle2022trend}.

The second mitigation strategy involves discarding some of the coarsest wavelet scales used in the periodogram bias correction (step 3 above). Instead of using all $J$ available scales as is customary in LSW modelling (see e.g., \cite{nason2000wavelet}), we instead use $J_0 <J$ scales. This can be thought of as a type of tapered estimator. Using $J_0 <J$ scales improves practical performance of the estimators in the presence of trends, and reduces the variability of the resulting estimator by discarding wavelet coefficients at coarser scales, which become increasingly autocorrelated.  \cite{mcgonigle2022trend} show the consistency of these estimators and demonstrate the superiority over the standard estimators of \cite{nason2000wavelet} in the presence of trends.

Thus the final TLSW estimator is given by
\begin{equation}\label{eq:spec-est1}
\widehat{S}_{j} (k/n) = \sum_{l=1}^{J_0}  A_{lj}^{-1} \widehat{I}_{l,k},
\end{equation}
and the LACV is estimated using $\widehat{S}_{j} (k/n)$ with the equation $\hat{c} (k/n, \tau) = \sum_{j=1}^{J_0} \widehat{S}_{j} (k/n) \Psi_{j} (\tau)$. % mix of negative and positive scales here, RK changed to be positive (same in next section too!)

\subsection{First-differenced estimation with modified correction factor}\label{sec:tlsw-spec2}

The first approach described above works well when the trend function does not display cusps. However, in the presence of such irregularities, the wavelet transform may fail to remove the trend function, causing the EWS estimator to be biased. In particular, wavelet coefficients at coarser scales will suffer from increased bias due to the increased filter length. Furthermore, since the periodogram must be corrected across scales, the bias at coarser scales can appear at finer scales.

To solve this problem, we can use the commonly used practice of differencing. Before applying the wavelet transform, we first difference the time series to yield $\{\Delta X_t  = X_{t+1} - X_{t}\}_{t=1}^{n-1}$. Differencing allows us to remove the trend immediately, ensuring that the wavelet transform does not accumulate large bias across increasing scales. Then, the wavelet coefficients $\tilde{d}_{j,k} = \sum_t \Delta X_{t} \psi_{j,k} (t)$ are computed on the differenced time series, from which the wavelet periodogram $\tilde{I}_{j,k} =  \tilde{d}_{j,k} ^2$ is constructed. \tcb{It is possible to use higher order differencing, e.g. second differencing, however we advise against this due to the inherent loss of information and poorer empirical performance as demonstrated in \cite{mcgonigle2022modelling}}.

Estimation then proceeds as before, with one key distinction. Due to the differencing transform, the dependence structure of the time series is changed. Therefore, the bias in the wavelet periodogram is no longer characterised by the matrix $A = (A_{jl})_{j, l >0}$, but instead a related matrix $D^1 = (D_{jl}^1)_{j, l >0}$. The entries of the modified bias matrix are given by $D_{jl}^1 = A_{jl} - A_{jl}^1$ where $A^{1}_{jl} = \sum_\tau \Psi_{j} (\tau) \Psi_l (\tau-1)$, and we have that
\begin{equation*}
\mathbb{E} (\tilde{I}_{k}^{j} ) \approx  \sum_{l} D_{jl}^{1} S_{l} \left( \frac{k}{n} \right).
\end{equation*}
Therefore, we can proceed similarly to before, first smoothing the periodogram by for example using a running mean:
\begin{equation*}
\widehat{\tilde{I}}_{l,k} = \frac{1}{2N+1} \sum_{m=-N}^{N} \tilde{I}_{j,k+m}, 
\end{equation*}
and then correcting using the matrix $(D^1)^{-1}$, giving the estimator
\begin{equation*}
\widehat{S}_{j} (k/n) = \sum_{l=1}^{J_0}  (D_{lj}^1)^{-1} \widehat{\tilde{I}}_{l,k}, 
\end{equation*}
where, as seen with the direct estimation approach, we opt to use a tapered estimator with maximum scale of interest $J_0 <J$. Analogously, for a more complex trend component, second differences can also be used. If the time series is known to have a seasonal trend of period $p$, then the spectrum can be instead estimated using the wavelet periodogram of the lag $p$ differenced time series:
\begin{equation}\label{eq:diff-spec}
\widehat{S}_{j} (k/n) = \sum_{l=1}^{J_0}  (D_{lj}^p)^{-1} \widehat{\tilde{I}}_{l,k}, 
\end{equation}
where $D_{jl}^p = A_{jl} - A_{jl}^p$ with $A^{p}_{jl} = \sum_\tau \Psi_{j} (\tau) \Psi_l (\tau-p)$.
For more details, we refer the reader to \cite{mcgonigle2022modelling}.  The local autocovariance can be estimated using this spectrum estimate in the same manner as the direct estimator, with $\hat{c} (k/n, \tau) = \sum_{j=1}^{J_0} \widehat{S}_{j} (k/n) \Psi_{j} (\tau )$.

\subsection{EWS estimation using TLSW}

In the \pkg{TrendLSW} package, the EWS estimation is performed as part of \texttt{TLSW()}, which will, by default, directly estimate the trend function and spectrum simultaneously. Here, we describe how the spectral estimation component of \texttt{TLSW()} is carried out.

To indicate to \texttt{TLSW()} that the user wishes to compute the spectral estimate, the argument \texttt{do.spec.est} should be set to \texttt{TRUE} (default). Then, the essential arguments to perform spectral estimation in \texttt{TLSW()} are:

\begin{itemize}
\item \texttt{x}: the input time series.
\item \texttt{S.filter.number}: the filter number of the wavelet used to estimate the EWS. By default this is $4$. Larger values are smoother, \texttt{"DaubExPhase"} can take values 1-10 and \texttt{"DaubLeAsymm"} values 4-10.
\item \texttt{S.family}: the family of the wavelet used to estimate the EWS. By default this is \texttt{"DaubExPhase"}, \texttt{"DaubLeAsymm"} is an alternative.
\item \texttt{S.smooth}: logical indicating whether to smooth, default \texttt{TRUE}.
\item \texttt{S.smooth.type}: the type of smoothing to be used on the wavelet periodogram. By default this is \texttt{"mean"} for running mean smoothing, but it can also be \texttt{"median"} for running median, or \texttt{"epan"} for Epanechnikov kernel smoothing.
\item \texttt{S.binwidth}:  the bin width parameter in the wavelet periodogram smoother; this was $2N+1$ in Equation~\ref{eq:runmean}. By default this is $\lfloor 6 \sqrt{n} \rfloor$.
\item \texttt{S.max.scale}: the maximum wavelet scale to be analysed, $J_0$ in Equation~\ref{eq:spec-est1}. By default this is $\lfloor 0.7 \log_2 n \rfloor$.
\item \texttt{S.boundary.handle}: a logical variable, with default value \texttt{TRUE}, indicating whether the boundary handling procedure is used.
\item \texttt{S.inv.mat}: The user can pre-calculate and supply a correction matrix to correct the raw wavelet periodogram. If left blank, then the correction matrix is calculated when performing spectral estimation.
\item \texttt{S.do.diff}: a logical variable, to indicate whether the time series should be differenced first before the wavelet periodogram is calculated. If \texttt{TRUE}, then the differencing approach is carried out from Section \ref{sec:tlsw-spec2}, otherwise the direct estimation approach is used from Section \ref{sec:tlsw-spec1}. By default this is \texttt{FALSE}.
\item \texttt{S.lag}: if \texttt{S.do.diff=TRUE}, the lag of the differencing operator to apply. The default choice is $1$, whilst larger values could be used if there is seasonality at a given lag (e.g., lag 12 could be used in monthly data).
\item \texttt{S.diff.number}: the number of times differencing is applied to the time series if the argument \texttt{S.do.diff=TRUE}. The strongly recommended default is $1$ to perform a first difference, but can be set to $2$ to perform second differencing as opposed to lag 2 differencing.
\end{itemize}

All arguments related to spectral estimation begin with the prefix \texttt{S}. By default, no differencing will be performed, so that the direct spectral estimation method from Section \ref{sec:tlsw-spec1} is performed. Default values for wavelet choice, maximum scale $J_0$, and bin width, have been set to align with numerical studies carried out in \cite{mcgonigle2022modelling}, \cite{mcgonigle2022trend}, and widely accepted in the literature.  All other arguments necessary for estimating the spectrum have been given default values based on extensive numerical studies. Therefore, basic estimation of the spectrum of the time series \texttt{x1} without manually adjusting tuning parameters can be performed by supplying \texttt{x1} to \texttt{TLSW()} as follows:

\begin{CodeChunk}
\begin{CodeInput}
R> x1.TLSW <- TLSW(x1)
\end{CodeInput}
\end{CodeChunk}

\texttt{TLSW()} returns an object with S3 class \texttt{TLSW}; a list which contains several of the quantities also returned by \texttt{ewspec3()} from the \texttt{locits} \proglang{R} package, as well as additional input parameters. Outputs associated to the spectrum estimation are stored in the \texttt{spec.est} element of the \texttt{TLSW} object, itself a list. In this list, the corrected, smoothed spectral estimate is stored in the \texttt{S} component, whilst the unsmoothed and smoothed wavelet periodograms are stored in the \texttt{WavPer} and \texttt{SmoothWavPer} components respectively. As in the \pkg{wavethresh} package, these three components are of class \texttt{wd} and can be plotted using the plotting functionality therein, as well as being amenable to \texttt{print()} or \texttt{summary()}.

A \texttt{TLSW} object can be plotted using \texttt{plot()} as standard: this can plot spectrum and/or trend estimates with a number of options available to the user. For now, we will focus on the produced spectrum plot, and return to more details of plot functionality in a worked example in Section \ref{sec:worked-examples}. Using \texttt{plot()}, we supply the \texttt{TLSW} object, and for now we specify the argument \texttt{plot.type = "spec"} to only plot the estimated spectrum:

\begin{CodeChunk}
\begin{CodeInput}
R> plot(x1.TLSW, plot.type = "spec")
\end{CodeInput}
\end{CodeChunk}

\begin{figure}[]
\centering
\includegraphics[width =0.9\textwidth]{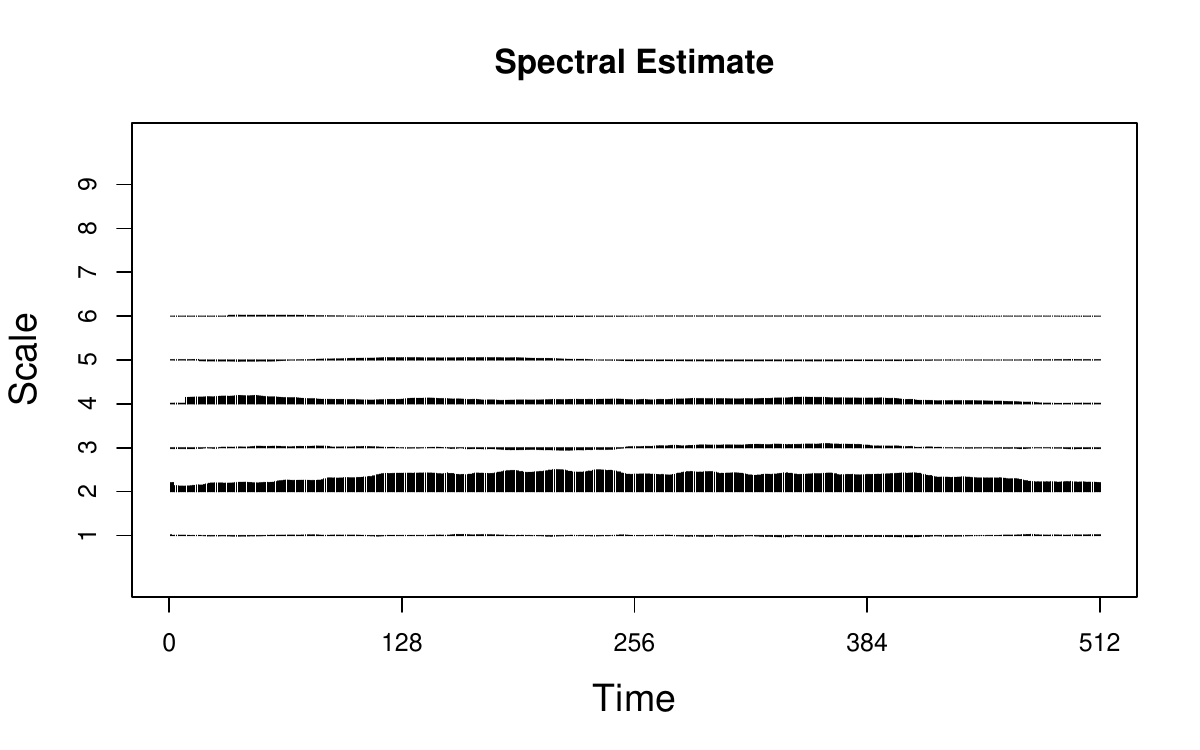}
\caption{Spectrum estimate for the TLSW process \texttt{x1} given by the (default) direct estimator using \texttt{TLSW()}.}
\label{fig:specex}
\end{figure}

The resulting spectrum estimate is shown in Figure~\ref{fig:specex}. The general features of the spectrum are well-represented in the estimate: scales with no power are generally estimated close to zero, whilst scale $4$ appears to have (near) constant power over time. The increasing and decreasing behaviour at scale 2 is also quite clear. Note that nonstationary spectrum estimation is an extremely challenging problem, due to factors including strong autocorrelation and low signal-to-noise ratio. Therefore, spectral estimates are not necessarily expected to be as accurate (as compared to the estimate of the trend function in Section~\ref{sec:trend-estimation}).

\begin{figure}[]
\centering
\includegraphics[width =0.9\textwidth]{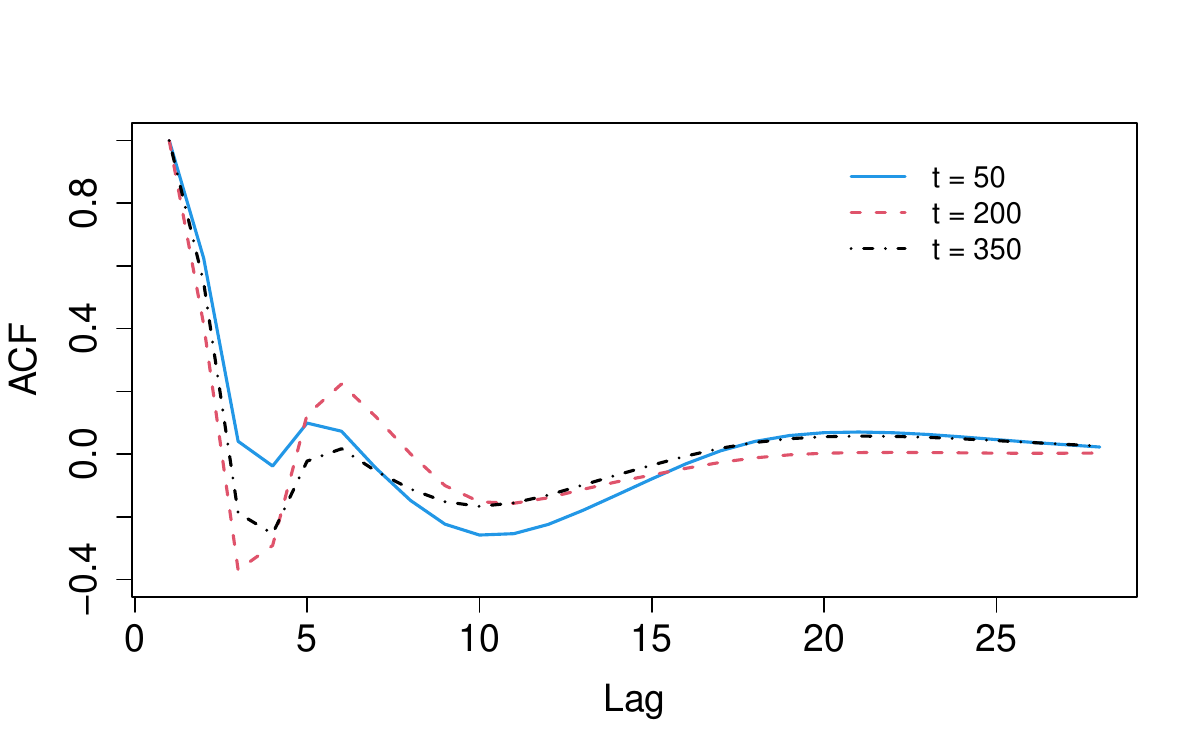}
\caption{Local autocorrelation estimate for the TLSW process \texttt{x1} at three time points $t = 50$ (blue solid line), $t = 200$ (red dashed line), and $t = 350$ (black dash-dotted line).}
\label{fig:lacf}
\end{figure}

The local autocovariance and autocorrelation of the times series can be computed from the spectral estimate using \texttt{TLSWlacf()}. The user can set a maximum lag for calculating the autocovariance in the same way as in \texttt{stats::acf()}, by specifying a value for the \texttt{lag.max} argument. \texttt{TLSWlacf()} takes as input a \texttt{TLSW} object produced from \texttt{TLSW()}, and outputs a list object of class \texttt{lacf}, which is identical to the output of \texttt{lacf()} in the \texttt{locits} package. The key outputs are \texttt{lacv} and \texttt{lacf}, which correspond to the estimated local autocovariance and autocorrelation respectively. These are given in matrix form with rows representing time and columns representing lags. We plot the autocorrelation estimate at time points $t = 50, 200, 350$ as an illustration in Figure~\ref{fig:lacf}.

\begin{CodeChunk}
\begin{CodeInput}
R> x1.lacf <- TLSWlacf(x1.TLSW)
R> plot(x1.lacf$lacr[50, ], ylim = c(-0.4, 1), type = "l", xlab = "Lag", 
        ylab = "ACF", lwd = 2, col = 4)
R> lines(x1.lacf$lacr[200, ], col = 2, lwd = 2, lty = 2)
R> lines(x1.lacf$lacr[350, ], col = 1, lwd = 2, lty = 4)
R> legend(20, 1, c("t = 10", "t = 230", "t = 450"), lty = c(1, 2, 4), 
          lwd = c(2, 2, 2), col = c(4, 2, 1), bty = "n", cex = 1.2)
\end{CodeInput}
\end{CodeChunk}

\section{Trend estimation}\label{sec:trend-estimation}

We now turn to describe two procedures for trend estimation in the presence of time-varying dependence structure. The general approach for trend estimation again consists of three main steps:

\begin{enumerate}
    \item Calculating the wavelet coefficients via a wavelet transform.
    \item Thresholding the wavelet coefficients.
    \item Obtaining the trend estimate by inverse transforming the thresholded wavelet coefficients.
\end{enumerate}

Here, the distinction between the two approaches lies in how the wavelet thresholding is performed. Mirroring section \ref{sec:ews-estimation}, the following sections discuss in detail how the two methods are implemented before exploring the package structure.

\subsection{Linear wavelet thresholding estimator}\label{sec:trend1}

In the case where the the trend function is sufficiently smooth, and the EWS was estimated using the direct approach described in Section \ref{sec:tlsw-spec1}, then we recommend to estimate the trend function using a linear wavelet thresholding estimator in the spirit of e.g., \cite{craigmile2004trend}. Observe that, when the trend is exactly a polynomial, using a wavelet with high enough vanishing moments will result in zero wavelet coefficients apart from at the boundaries; the zero mean assumption for the the LSW increment process in (\ref{lsw_rep}) means that the wavelet coefficients have expectation zero. Thus the boundary wavelet coefficients and the scaling coefficients will be the only coefficients to meaningfully contribute to the trend component. Likewise, if the trend is closely approximated by a polynomial, the trend component is well-represented by only the boundary and scaling coefficients. 

We hence proceed as follows. First we perform a discrete wavelet transform (DWT) of the time series with a given choice of wavelet $\psi$ and a pre-specified coarsest scale $J_0$, similar to the spectral estimation procedure. Setting the non-boundary coefficients to zero will reflect the above observation on these coefficients being zero mean, in effect performing hard thresholding of the non-boundary coefficients, so that performing the inverse discrete wavelet transform will obtain a trend estimate.  \cite{mcgonigle2022trend} show that this procedure results in an unbiased and mean square consistent estimator of the trend when the true underlying trend is polynomial or H\"older continuous. Experiments have shown that in practice, setting a coarsest scale of $J_0 = \lfloor 0.7\log_2(T) \rfloor$ is robust to a wide range of process scenarios.

To obtain robust estimation of the trend at the boundary of the series, \cite{mcgonigle2022trend} suggest using a procedure akin to classical boundary handling in wavelet methods, in which a long series of length $3n$ is constructed by reflecting the time series data so that the original series is at the centre of the newly constructed data.  Performing trend estimation on this longer series mitigates the fact that, particularly at coarser scales, the wavelet filter can be longer than the available data near the boundary.  Note that a nondecimated discrete wavelet transform (NDWT) can be performed instead of the DWT to represent more temporally-localised behaviour in the series, in which case basis averaging is used to obtain the trend estimate after hard thresholding, see e.g., \citet[Chapter 3.12]{nason2000wavelet} for more details. \tcb{The key difference between the DWT and NDWT is that DWT forms an orthogonal basis whereas the NDWT is an overcomplete basis representation.  The DWT downsamples the data at each step of the algorithm to obtain a sparse (efficient) representation of the underlying signal.  The disadvantage of this is that if the data is shifted by a single place to the left or right, the DWT representation can change dramatically (not just shifted one to the left/right).  In contrast, the NWDT is translation-invariant as it considers all potential paths of shifting and downsampling the data at each step of the algorithm and does not suffer from this drawback.}

When the DWT is used, pointwise $(1-\alpha)$\% confidence intervals for the trend can be derived under the assumption of normality, via
\begin{equation}\label{eq:linCI}
\widehat{T}_t\pm q_{1-\tfrac{\alpha}{2}}\sqrt{\operatorname{Var} \left( \widehat{T}_t \right)},
\end{equation}
where the variance term can be estimated using the (consistent) estimate of the local autocovariance $\hat{c} (k/n, \tau) = \sum_{j=-J_{0}}^{-1} \widehat{S}_{j} (k/n) \Psi_{j} (\tau )$.

When the NDWT is used, pointwise confidence intervals are computed via the bootstrap, in a similar fashion to the approach of \cite{friedrich2020autoregressive}. For some chosen total number of bootstrap replications, $B$, bootstrapped replications $\{X_t^{(b)} \}_{t=0}^{n-1}$ for $b = 1, \ldots, B$ can be simulated using the spectral estimate and trend estimate, from which bootstrapped trend estimates can be calculated. 

Concretely, using the spectral estimate $\{ \widehat{S}_j (k/n) \}_{j=1}^{J_0}$, for $k=0, \ldots, n-1$, and the trend estimate $\widehat{T}_t$, bootstrapped replications
\begin{equation}
X^{(b)}_t = \widehat{T}_t + \sum_j \sum_k \widehat{w}_{j,k} \psi_{j,t-k} \xi^{\tcb{(b)}}_{j,k}    
\end{equation}
are generated, where $\widehat{w}_{j,k} = \widehat{S}_j (k/n)^{1/2}$, and $\xi^{\tcb{(b)}}_{j,k}$ are independent, identically distributed standard Normal random variables. For each of the $B$ bootstrapped replications, a bootstrapped trend estimate \tcb{$\{ \widehat{T}_t^{(b)}\}_{t=0}^{n-1}$} is obtained using the linear wavelet thresholding procedure as described above. Then, for each time point $t = 0, \ldots, n-1$, pointwise $(1-\alpha)$\% confidence intervals can be calculated using either:

\begin{enumerate}
    \item a normal approximation $\widehat{T}_t\pm q_{1-\tfrac{\alpha}{2}}\sqrt{\operatorname{Var} ( \widehat{T}_t )}$, where the variance term is the sample variance of the $B$ bootstrapped trend estimates at time $t$.
    \item the empirical $\tfrac{\alpha}{2}$\% and $(1-\tfrac{\alpha}{2})$\% quantiles of the $B$ bootstrapped trend estimates.
\end{enumerate}

\subsection{Nonlinear wavelet thresholding estimator}\label{sec:trend2}

In the case where the EWS is estimated using the differencing approach of \cite{mcgonigle2022modelling} described in Section \ref{sec:ews}, then we recommend to estimate the trend function using a nonlinear wavelet thresholding estimator, using the spectral estimate $\widehat{S}_{j} (k/n)$ from Equation~\ref{eq:diff-spec}. More specifically, we employ classical wavelet thresholding to provide a trend estimate.  For a particular wavelet basis used for thresholding, $\{\psi^1_{j,k}\}_{j,k} $, we first compute the corresponding DWT wavelet coefficients of the series $X_t$, $d^1_{r,s} =  \sum_{t} X_t \psi^1_{j,k} (t)$, where we use the notation $d^1_{r,s}$ to differentiate from the wavelet coefficients corresponding to the generating wavelet in model \ref{eq:model1}. The coefficients are then thresholded using a coefficient-specific hard or soft threshold 
\begin{eqnarray*}
\hat{v}_{r,s}^S&=&\operatorname{sgn}(d^1_{r,s})(|d^1_{r,s}|-\lambda_{r,s,n})\mathbb{I}(|d^1_{r,s}| > \lambda_{r,s,n}) \qquad \textrm{(soft thresholding)}\\
\hat{v}_{r,s}^H&=&d^1_{r,s}\mathbb{I}(|d^1_{r,s}| > \lambda_{r,s,n}), \hspace{3.8cm} \textrm{(hard thresholding)}
\end{eqnarray*}
where $\lambda(r,s,n)=\hat{\sigma}_{r,s}\sqrt{2\log(n)}$ and $\hat{\sigma}_{r,s}$ is an estimate of the standard deviation of the coefficient $d^1_{r,s}$.  This estimate can be obtained from the expression for the variance of the wavelet coefficients 
\begin{equation}\label{eq:threshvar}
    \operatorname{Var}(d^1_{r,s}) \approx \sum_{l} C^{1,0}_{r,l}S_{l}(s/n),
    \end{equation}
where $C^{1,0}_{r,l} = \sum_\tau \Psi^0_{r} (\tau) \Psi^1_l (\tau)$ is computed using the autocorrelation wavelets $\Psi^0_{r} (\tau)$ and $\Psi^1_l (\tau)$ corresponding to the generating wavelet $\psi$ and thresholding wavelet $\psi^1$ respectively  \citep{mcgonigle2022modelling}.  The estimate and therefore the coefficient-dependent threshold $\lambda(r,s,n)$ is computed by plugging in the spectral estimate $\widehat{S}_{j} (k/n)$ from Equation~\ref{eq:diff-spec} into Equation~\ref{eq:threshvar}; a trend estimate $\widehat{T}_t$ can then be obtained by performing the inverse DWT on the coefficients $\{\hat{v}_{r,s}\}_{r,s}$.  

As in the linear wavelet thresholding approach, a pre-specified coarsest scale $J_0$ is set for the wavelet decomposition. Stronger estimation performance is also achieved using so-called translation-invariant (TI) denoising, in which the thresholding procedure described here is performed on the coefficients resulting from an NDWT on the data, followed by basis averaging; see \cite{von2000non} or \cite{mcgonigle2022modelling}.  Boundary handling can also be performed, in a similar manner to that described for the linear wavelet estimator. Lastly, pointwise $(1-\alpha)$\% confidence intervals can be construted via bootstrapping in an analogous way to that described for the linear wavelet estimator.

\subsection{Trend estimation using TLSW}

Both trend estimation procedures described above can be performed using \texttt{TLSW()}, setting the \texttt{do.trend.est} argument to \texttt{TRUE} (default).  By default, the function will use the linear wavelet estimator (paired with direct spectrum estimation), but the nonlinear (translation-invariant) wavelet-based estimation can be chosen by setting the \texttt{T.est.type} argument to \texttt{"nonlinear"}.

For both linear and nonlinear wavelet trend estimation, the following arguments can be specified by the user:

\begin{itemize}
\item \texttt{T.filter.number}: the filter number of the wavelet used to estimate the trend. By default this is $4$.
\item \texttt{T.family}: the family of the wavelet used to estimate the trend. By default this is \texttt{"DaubExPhase"}.
\item \texttt{T.transform}: the type of wavelet transform used in the thresholding estimator, either \texttt{"dec"} (DWT) or \texttt{"nondec"} (NDWT, default).
\item \texttt{T.boundary.handle}: a logical variable, with default value \texttt{TRUE}, indicating whether the reflection-based boundary handling procedure should be performed.
\item \texttt{T.max.scale}: the maximum wavelet scale to be used in the wavelet transform, $J_0$. By default this is $\lfloor 0.7 \log_2 n \rfloor$.
\item \texttt{T.CI}: a logical variable, to indicate whether a pointwise confidence interval for the trend should be computed. By default this is \texttt{FALSE}.
\item \texttt{T.sig.lvl}: a significance level, $\alpha$ for the constructed confidence interval, defaulting to \texttt{0.05}.
\item \texttt{T.reps}: the total number of bootstraps, $B$, used in the construction of pointwise confidence intervals, if \texttt{T.CI} is \texttt{TRUE}, defaulting to $B=200$.
\item \texttt{T.CI.type}: the type of constructed confidence intervals, if \texttt{T.CI} is \texttt{TRUE}. By default this is \texttt{"normal}, which uses the normal approximation quantiles to construct the interval, else \texttt{"percentile"} for which empirical quantiles are used in the construction.
\item \texttt{T.lacf.max.lag}: The maximum lag used for calculating the lacf, default is $\lfloor 10\log n \rfloor$,
\end{itemize}

For the nonlinear trend estimation, as well as the arguments above, the user can specify the following additional arguments associated to how the wavelet thresholding is performed:

\begin{itemize}
\item \texttt{T.thresh.type}: the type of thresholding used in the nonlinear quantiles trend estimation procedure, either \texttt{"hard"} (default) or \texttt{"soft"}.
\item \texttt{T.thresh.normal}: a logical variable, indicating whether to use a larger threshold, useful to disable for non-Normally distributed data. Defaults to \texttt{TRUE}, in which case $\lambda(r,s,n)=\hat{\sigma}_{r,s}\sqrt{2\log(n)}$ is used, if \texttt{FALSE}, then $\lambda(r,s,n)=\hat{\sigma}_{r,s} \log(n)$ is used.
\end{itemize}

In a similar fashion to that described in Section~\ref{sec:ews-estimation}, we can use \texttt{plot()} on the \texttt{x1.TLSW} object to plot the trend estimate, by supplying the argument \texttt{plot.type = "trend"} only the trend is plotted:

\begin{CodeChunk}
\begin{CodeInput}
R> plot(x1.TLSW, plot.type = "trend")
\end{CodeInput}
\end{CodeChunk}

\begin{figure}[H]
\centering
\includegraphics[width =0.9\textwidth]{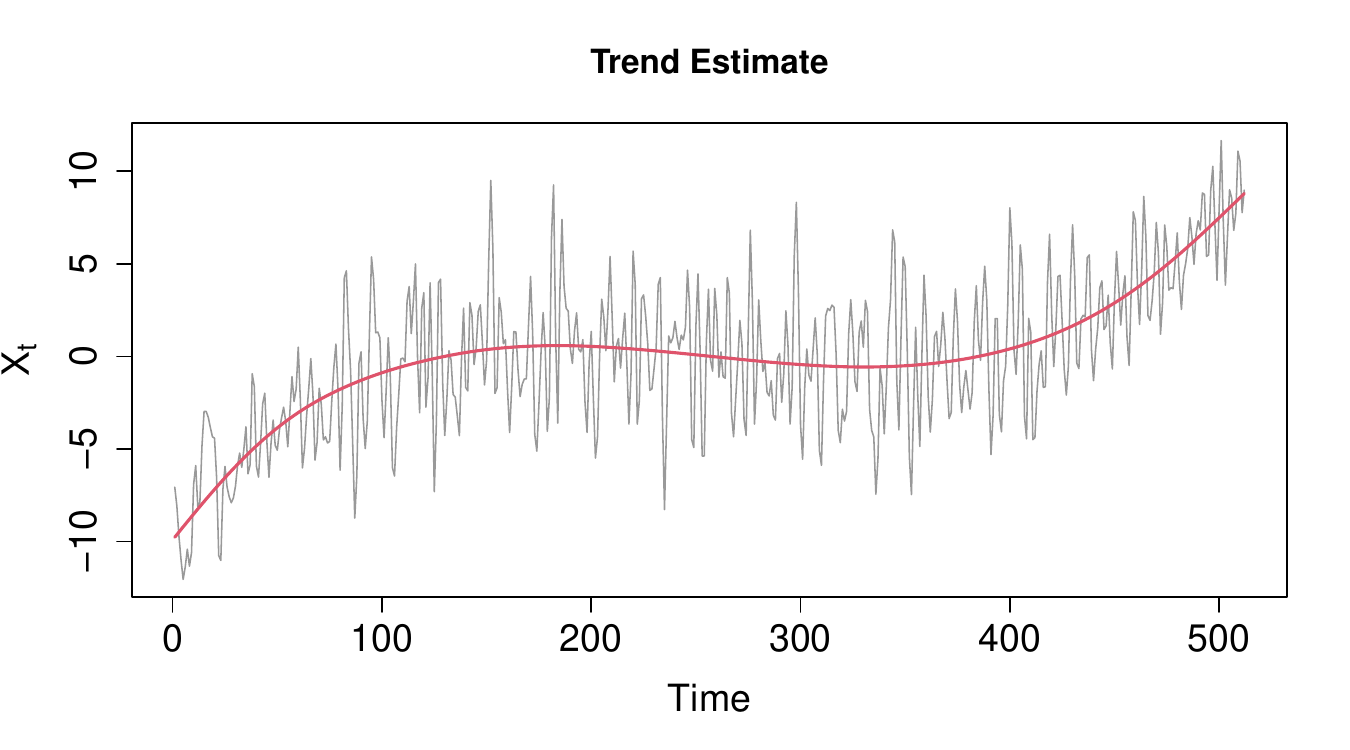}
\caption{Trend estimate (red) for the TLSW process \texttt{x1} given by the linear wavelet estimator described in the text using \texttt{TLSW()}. Underlying data given by grey line.}
\label{fig:trendex}
\end{figure}

The resulting plot is shown in Figure~\ref{fig:trendex}, which shows close alignment between the estimated and true trend function.

\section{Trend and EWS worked example}\label{sec:worked-examples}

Having described the main functionality of the package, in this section we give an extended worked example combining both trend and EWS estimation. We simulate a realisation of a TLSW process with trend function given by:
\begin{equation*}
T_t = \begin{cases} 5 \sin (6 \pi t/n) + 10t/300, &  0 \leq t  \leq 300, \\
                    5 \sin (6 \pi t/n) -4 - 14t/(n-300) - 14n/(300-n) , & 301 \leq t  < n .
 \end{cases}
\end{equation*}
i.e., a composition of a sinusoidal component, and a piecewise linear trend starting at $0$, increasing to 10 at time point $300$, then decreasing to $-4$ at time point $n$.
The spectrum is given by:
\begin{equation*}
S_j (t/n) = \begin{cases}  8t/n + 2, & j = 1, \quad 0 \leq t  < n, \\
                    1 , & j = 3, \quad 0 \leq t  \leq 200 \\
                    (5/200)t - 4 , & j = 3, \quad 200 \leq t  \leq 400 \\
                    -(5/200)t  +16 , & j = 3, \quad 400 \leq t  < 600 \\
                    1 , & j = 3, \quad 600 \leq t  < n \\
                    2 + 4 \sin (4 \pi t/n)^2, & j = 5, \quad 0 \leq t  < n.
 \end{cases}
\end{equation*}
i.e., a linearly increasing power at scale $j=1$, piecewise linear with peak in power at time point $400$ at scale $3$, and sinusoidal power at scale $j=5$. This time, we specify the trend function in \proglang{R} using a numeric vector, and specify the spectrum using a matrix of numeric values, as follows:
\begin{CodeChunk}
\begin{CodeInput}
R> n2 <- 1024
R> index2 <- seq(from = 0, to = 1, length = n2) 
R> trend2 <- 5 * sin(pi * 6 * index2) + 
             c(seq(from = 0, to = 10, length = 300),  
             seq(from = 10, to = -4, length = 724))
R> spec2 <- matrix(0, nrow = log2(n2), ncol = n2)
R> spec2[1,] <- seq(from = 2, to = 10, length = n2)
R> spec2[3,] <- c(rep(1, 200), seq(from = 1, to = 6, length = 200),
                  seq(from = 6, to = 1, length = 200), rep(1, 424))
R> spec2[5,] <- 2 + 4 * sin(4 * pi * index2)^2
\end{CodeInput}
\end{CodeChunk}
As before, we simulate the TLSW process via \texttt{TLSWsim()}:
\begin{CodeChunk}
\begin{CodeInput}
R> set.seed(1234)
R> x2 <- TLSWsim(trend = trend2, spec = spec2)
\end{CodeInput}
\end{CodeChunk}

\begin{figure}[]
\centering
\includegraphics[width =0.9\textwidth]{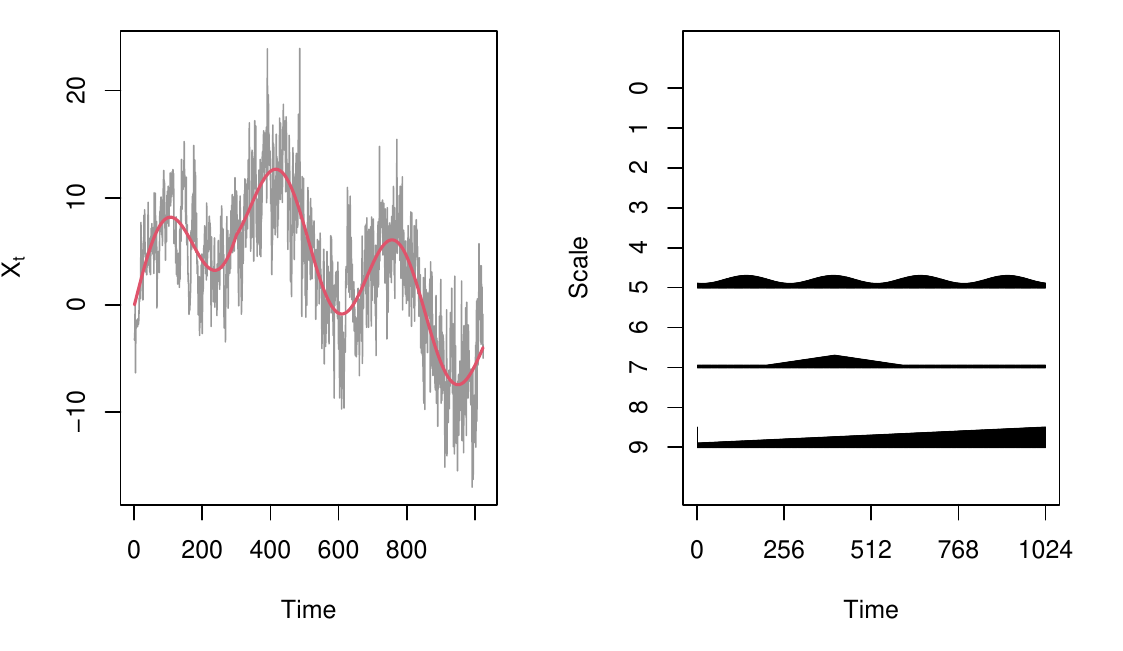}
\caption{Left: plot of an example realisation of the TLSW process \texttt{x2}, with true trend line given in red solid line. Right: true underlying spectrum.}
\label{fig:example2}
\end{figure}

In Figure~\ref{fig:example2} we plot \texttt{x2}, along with the true underlying trend function and spectrum.

%RK: need the code for the figure in replication material.

For estimating the trend and spectrum of this TLSW process, we make some modifications to the default arguments of \texttt{TLSW()} to highlight some potential ways a user can customise the estimation procedure. The following code snippet fits the TLSW model to \texttt{x2} shown in Figure~\ref{fig:example2}:
\begin{CodeChunk}
\begin{CodeInput}
R> set.seed(10)
R> x2.TLSW <- TLSW(x2, T.filter.number = 6, T.family = "DaubLeAsymm",
                T.est.type = "nonlinear", T.CI = TRUE,  T.reps = 500,
                S.do.diff = TRUE, S.smooth.type = "median", 
                S.binwidth = 128)
\end{CodeInput}
\end{CodeChunk}
In this example, we perform trend estimation using nonlinear thresholding, and use the Daubechies Least Asymmetric wavelet with 6 vanishing moments for trend estimation. The choice of wavelet is an important consideration in wavelet analysis, and is akin to selecting the kernel in nonparametric modelling. It is recommended to check the robustness of any conclusions to the choice of wavelet: the best performing wavelet will depend on the unknown trend function. A point-wise confidence interval is calculated for the trend estimate (which by default is at the $95$\% significance level) using $500$ bootstrap replications. 

\begin{figure}[]
\centering
\includegraphics[width =0.9\textwidth]{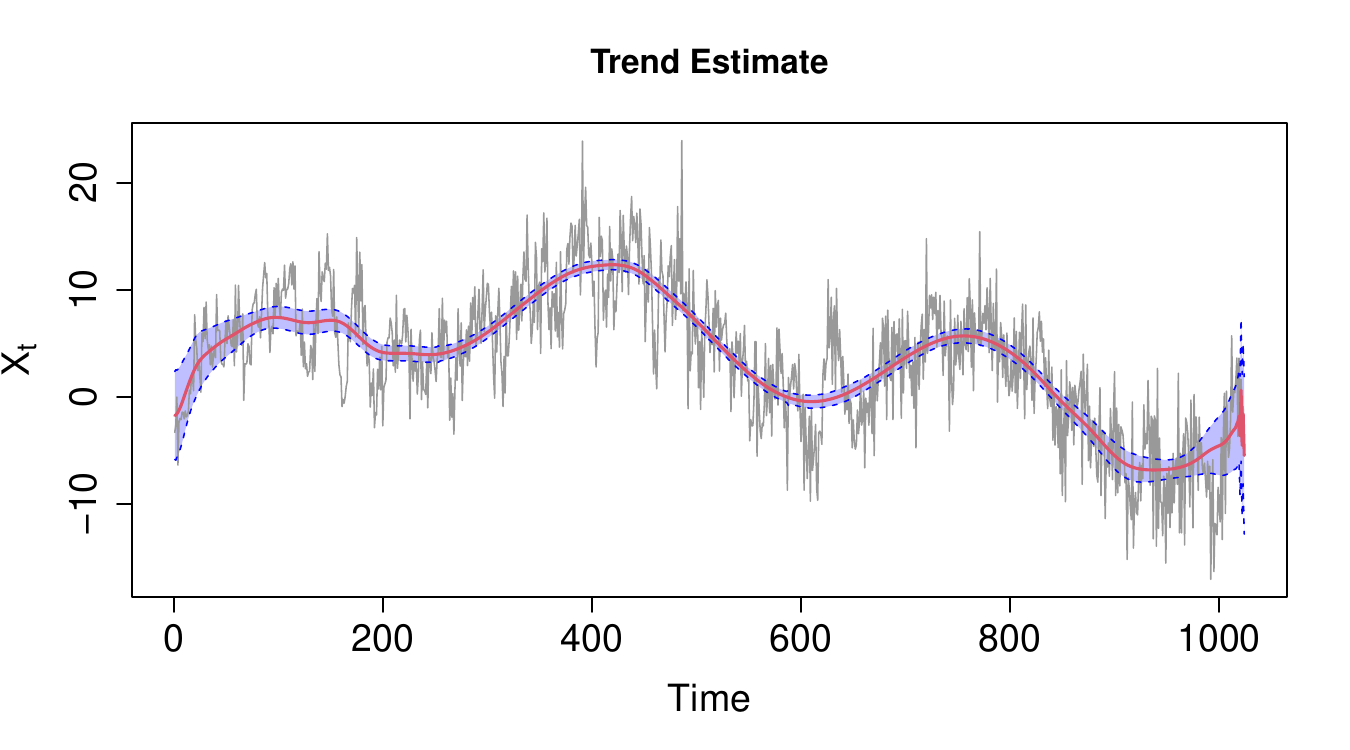}
\caption{Estimated trend function (red solid line) with 95\% pointwise confidence interval given by the blue shaded region and dashed blue lines. Original time series shown in grey solid line.}
\label{fig:example2-trend-est}
\end{figure}

The resulting estimate is shown in Figure~\ref{fig:example2-trend-est}, which aligns closely with the true trend function. By default, the calculated confidence interval is plotted, but this can be changed by setting the argument \texttt{plot.CI = FALSE} when calling \texttt{plot()}. As expected, the confidence intervals tend to be wider at the boundaries due to the boundary handling procedure and boundary effects.

\begin{figure}[]
\centering
\includegraphics[width =0.9\textwidth]{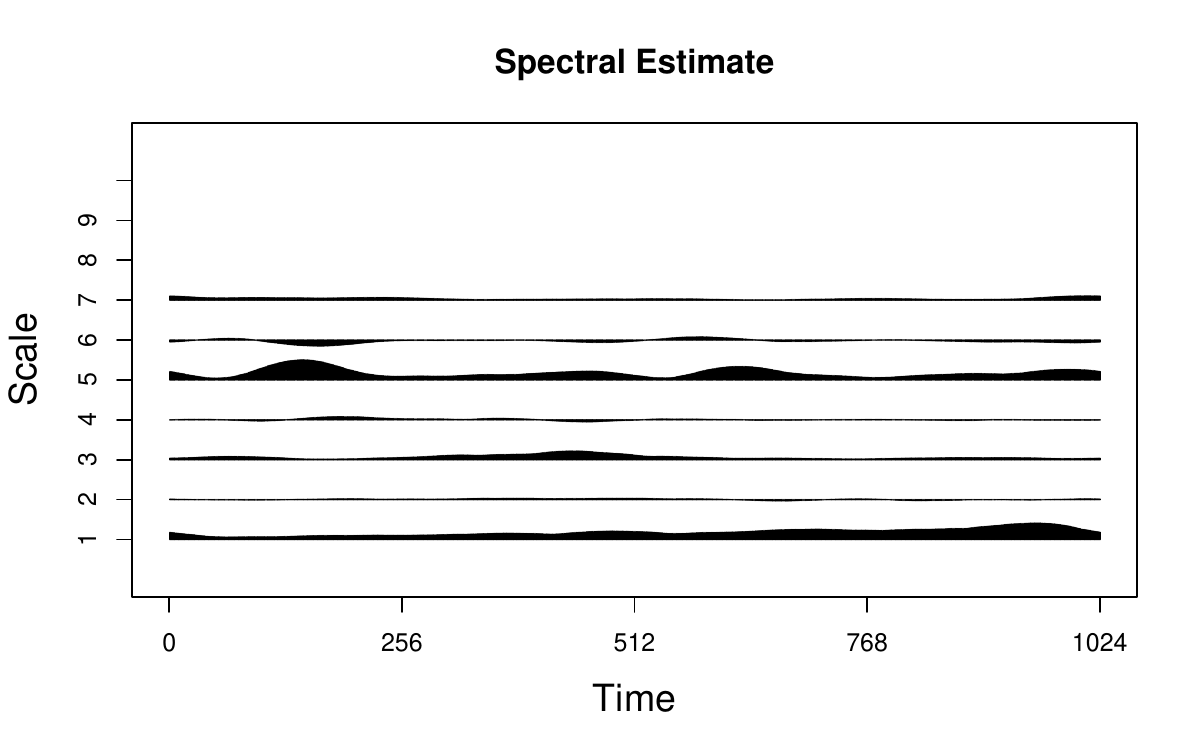}
\caption{Estimated spectrum. Scaling is applied globally across each scale for direct comparison.}
\label{fig:example2-spec-est}
\end{figure}

The spectrum estimate is computed using the differenced time series, (paired with the nonlinear trend estimate) and smoothed with a running median of bin width size $128$. A running median can be used for additional robustness in the estimator; for further information see \cite{mcgonigle2021detecting}. A smaller bin width is used due to the quickly evolving spectral structure. The estimate is shown in Figure~\ref{fig:example2-spec-est}. The estimate correlates well with the general features of the underlying spectrum, such as the slowly increasing power at scale 1, bump at scale 3, and the periodic pattern at scale 5.

\subsection{Making modifications to spectrum and trend estimation plots}

In Sections~\ref{sec:ews-estimation} --~\ref{sec:worked-examples} above, we used default settings to produce plots of the spectrum and trend estimates with the \pkg{TrendLSW} package.  However, the underlying method \texttt{plot.TLSW()} has flexible functionality to allow practitioners to create user-controlled plots of the estimated quantities of interest from the TLSW model.  More specifically, the style of the plots is controlled by the following arguments:

\begin{itemize}
\item \texttt{trend.plot.args}: a list object, giving options to modify the plotting of the TLSW trend estimate, with structure specified as follows:
\begin{itemize}
    \item Graphical parameters related to the display of the overall plot are inherited from \texttt{plot.default} and are specified in the usual way: for example, to change the title of the plot to ``Plot", use \texttt{main = "Plot"}.
    \item Parameters affecting the display of the estimated trend line should begin with the prefix ``\texttt{T.}". For example, to set the colour of the trend line to blue, use \texttt{T.col = "blue"}.
    \item Parameters affecting the display of the confidence interval lines should begin with the prefix ``\texttt{CI.}". For example, to set the line width of the confidence interval to 2, use \texttt{CI.lwd = 2}.
    \item Parameters affecting the display of the polygon drawn by the confidence interval should begin with the prefix ``\texttt{poly.}". For example, to set the colour of the confidence interval region to green, use \texttt{poly.col = "green"}.
\end{itemize}
\item \texttt{spec.plot.args}: a list object giving options to modify a plot of the TLSW spectrum estimate, for example axes specification, inherited from \texttt{plot.wd}.
\item \texttt{plot.CI}: a logical variable indicating whether the confidence interval of the trend estimate (if calculated) should be displayed.
\item \texttt{...}: any additional parameters that will modify both the trend and spectrum plots, e.g., \texttt{cex.main = 2}.
\end{itemize}

We now give an example of this plotting functionality for the \texttt{x2} example. The code below plots both the trend and spectral estimate as in the example from Section 3.3, with modifications to the graphical parameters of the plot.

\begin{CodeChunk}
\begin{CodeInput}
R> plot(x2.TLSW, trend.plot.args = list(col = "black", type = "p", pch = 16, 
                            T.col = "blue", T.lwd = 2, poly.col = "grey",
                            CI.col = "purple", CI.lwd = 2, CI.lty = 1),
        spec.plot.args = list(scaling = "by.level", ylab = "Level"))
\end{CodeInput}
\end{CodeChunk}

\begin{figure}[]
\centering
\includegraphics[width = 0.9\textwidth]{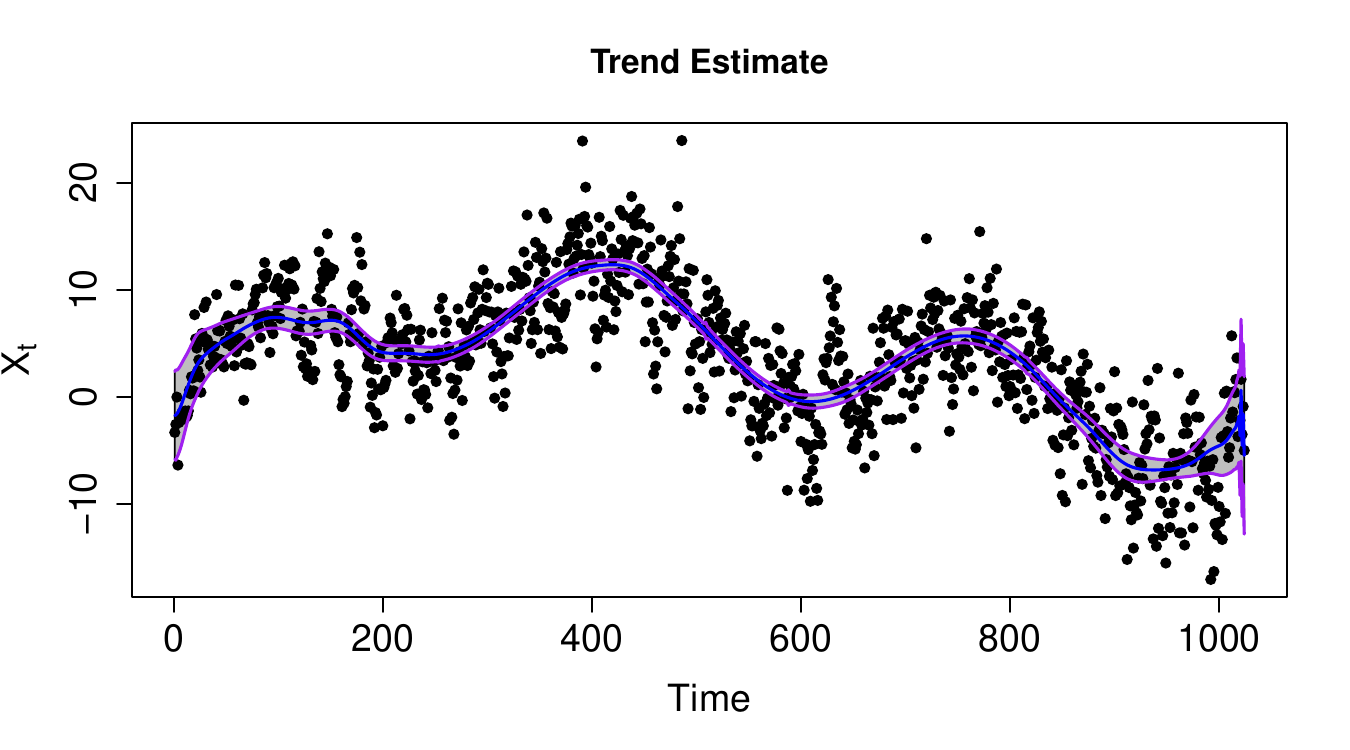}
\caption{Plot of the trend estimate for the TLSW process \texttt{x2} given by the code snippet in the text. }
\label{fig:example2-trend-mod}
\end{figure}

This produces a modified plot of the trend and spectrum for the \texttt{x2} series. The features of the trend estimate plot are changed by supplying options to the \texttt{trend.plot.args} argument in the form of a list. The underlying data in the trend plot have been changed to be black circles, by setting \texttt{col = "black", type = "p", pch = 16} in the \texttt{trend.plot.args} argument. The colour and line width of the trend estimate are set using \texttt{T.col = "blue", T.lwd = 2} inside \texttt{trend.plot.args}. The colour of the confidence interval's shaded region is set using \texttt{poly.col = "grey"}, and the features of the confidence interval lines are changes using \texttt{CI.col = "purple", CI.lwd = 2, CI.lty = 1}. The resulting trend plot is shown in Figure~\ref{fig:example2-trend-mod}.

\begin{figure}[]
\centering
\includegraphics[width =0.9\textwidth]{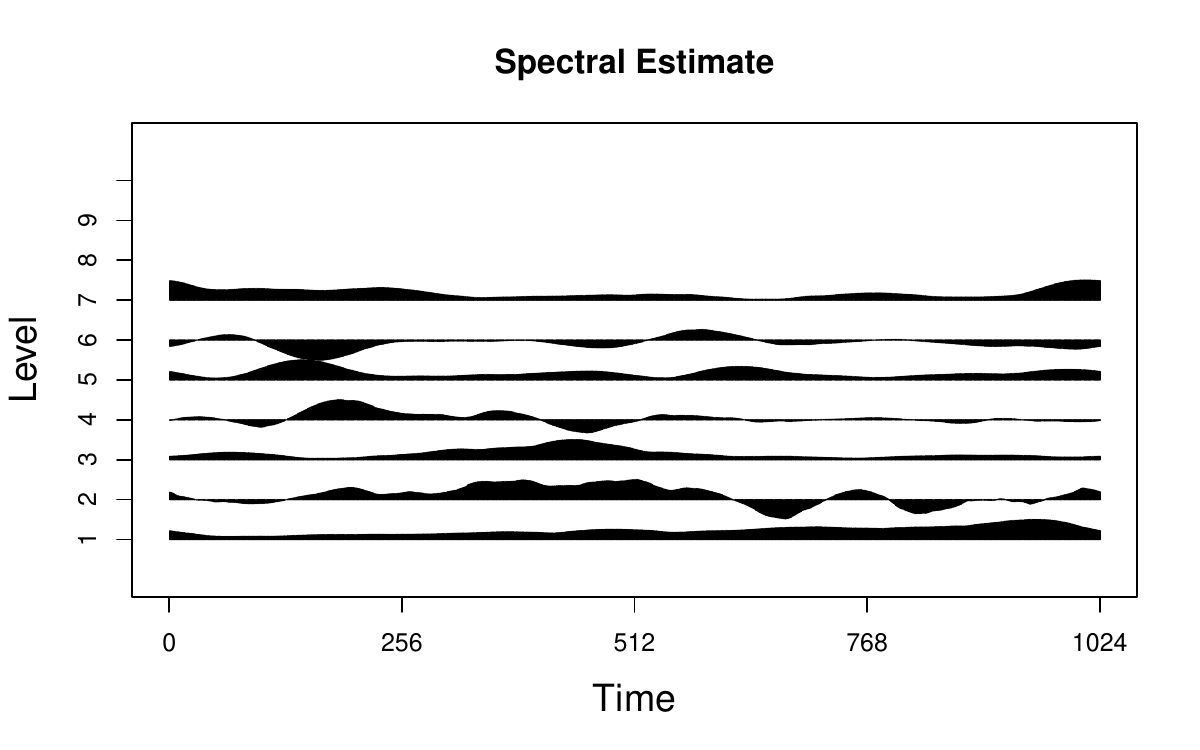}
\caption{Plot of the spectrum estimate for the TLSW process \texttt{x2} given by the code snippet in the text. Each level is scaled individually.}
\label{fig:example2-spec-mod}
\end{figure}

Features of the spectrum estimate plot can be changed by supplying options to the argument\texttt{spec.plot.args}: the scaling on each level of the plot (an option that is inherited from \texttt{wavethresh::plot.wd()}) is set to be individually scaled by specifying \texttt{scaling = "by.level"}, and the y-axis label has been modified using \texttt{ylab = "Level"}. The spectrum plot is shown in Figure~\ref{fig:example2-spec-mod}.

\section{Data examples}\label{sec:data-example}

We now describe the usage of the various methods described in Sections~\ref{sec:ews-estimation} and~\ref{sec:trend-estimation}  to analyse two real data examples.

\subsection{Bioluminescence of Caenorhabditis elegans}
\begin{figure}
    \begin{center}
        \includegraphics[width =0.8\textwidth]{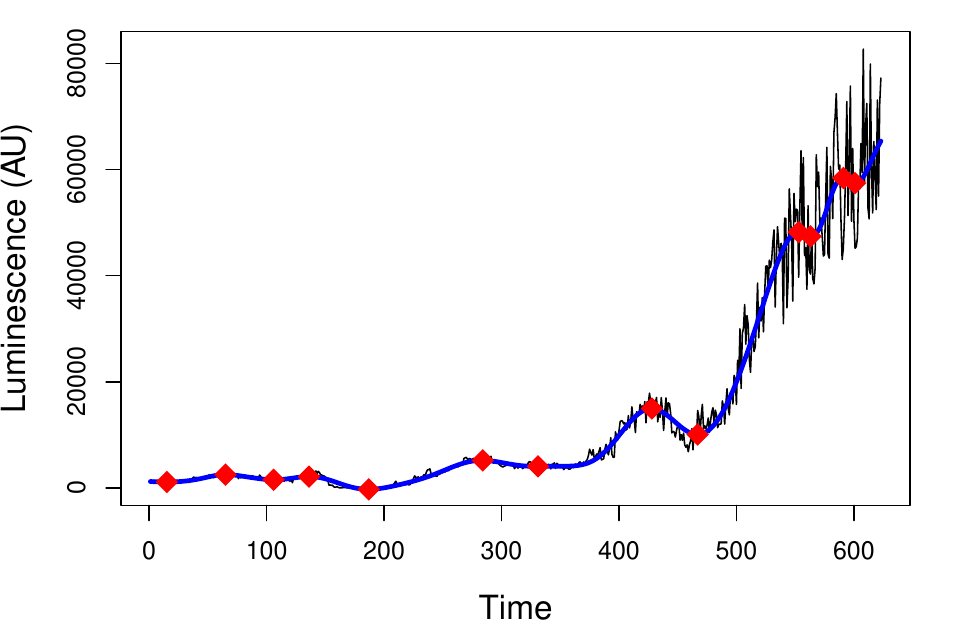}        
    \end{center}
    \caption{Bioluminescence of a group of C. Elegans peaks and troughs correspond to group feeding and growing phases respectively. Black thin line is the observed data, blue thick line is the TLSW trend fit and the red diamonds are the identified peaks and troughs from the TLSW trend fit.}\label{fig:bio}
\end{figure}

%RK: this has been checked and text added by Alex
Caenorhabditis elegans is an excellent model for high-throughput experimental approaches \citep{worms2}.  Notably, C. elegans transgenic strains have been develop expressing the firefly luciferase enzyme to produce bioluminescence to track the worm metabolic activity over time \citep{worms1}.  This has been applied to studying compound toxicity \citep{worms3}, ageing \citep{worms4} and development \citep{worms5}. During its life cycle, C. elegans goes through four successive larval stages and molts, during which they temporarily stop feeding and exhibit lower metabolic activity. This can be revealed by bioluminescence readings when the transgenic worms aforementioned are grown in presence of the luciferase substrate luciferin, leading to multimodal bioluminescence readings (Figure~\ref{fig:bio}) that are hard to fit with traditional functions, to automatically extract key parameters \citep{worms5}.

While manual determination of developmental phases and metabolic activity can be achieved easily, such assays are typically carried out in multiple batches of 96 or 384 well arrays, for which manual analyses would take far longer than the experiments themselves. It is challenging for bioscientists to automatically process bioluminescence traces due to the clear non-stationary second-order behaviour present in the time series. Transitions between molts and growth phases produce peaks and troughs of varying amplitude and width in these bioluminescence time series. Here we seek to measure these by identifying the peaks and troughs in the estimated trend whilst accounting for the time varying second order properties. The C. elegans data presented in Figure~\ref{fig:bio} is a time series of bioluminescence readings from 20 L1 worms (strain PE255) taken every six minutes over a 54h period in a M200 Infinite Pro Tecan plate-reader captured by Alexandre Benedetto (Lancaster University). The time series is available as \texttt{celegensbio} within the \pkg{TrendLSW} package. This is one time series taken from an experiment where hundreds of time series are collected \citep{worms3}.

The following code performs TLSW trend estimation on the C. Elegans time series and identifies the peaks and troughs within the estimated trend to produce Figure~\ref{fig:bio}.

\begin{CodeChunk}
\begin{CodeInput}
R> data("celegensbio")
R> out <- TLSW(celegensbio, T.filter.number = 10, T.family = 'DaubExPhase')
R> plot(out, plot.type = "trend", 
        trend.plot.args = list(ylab = "Luminescence (AU)", T.col = "blue", 
        T.lwd = 3, main = ""), cex.lab = 1.3, col = "black")
R> delta <- diff(out$trend.est$T)
R> turns <- which(delta[-1] * delta[-length(delta)] < 0) + 1
R> points(turns, out$trend.est$T[turns], col = "red", pch = 18, cex = 2)

\end{CodeInput}
\end{CodeChunk}

\tcb{It can be beneficial to fit a range of wavelets and assess the visual fit to the data in order to choose the most appropriate wavelet for trend estimation. In the code below, we compute the trend estimate for a range of wavelets.}

\begin{CodeChunk}
\begin{CodeInput}
R> out.EP1 <- TLSW(celegensbio, T.filter.number = 1, 
                    T.family = "DaubExPhase", T.transform = "nondec")
R> out.EP4 <- TLSW(celegensbio, T.filter.number = 4, 
                   T.family = "DaubExPhase", T.transform = "nondec")
R> out.LA10 <- TLSW(celegensbio, T.filter.number = 10, 
                    T.family = "DaubLeAsymm", T.transform = "nondec")
\end{CodeInput}
\end{CodeChunk}                    

\begin{figure}
    \begin{center}
        \includegraphics[width =0.8\textwidth]{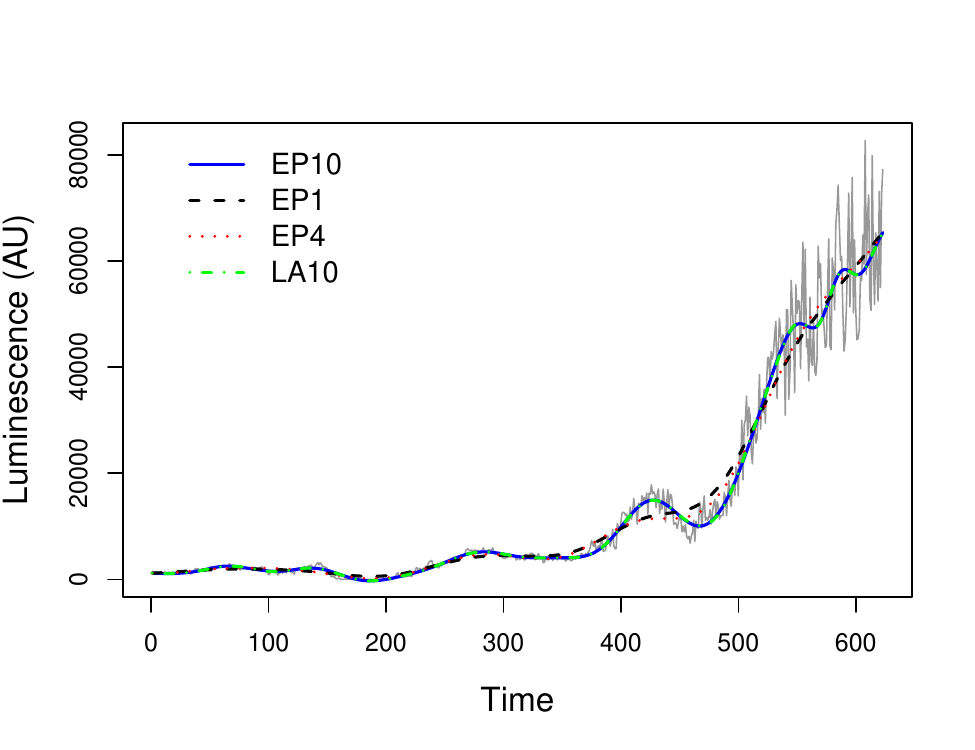}        
    \end{center}
    \caption{Comparison of TLSW trend fits obtained using: the EP10 wavelet (blue solid line), EP1 wavelet (black dashed line), EP4 wavelet (red dotted line), LA10 wavelet (green dot-dashed line).}
    \label{fig:bio-trend}
\end{figure}

\tcb{We can then visually compare the fitted trends, as shown in Figure~\ref{fig:bio-trend}.}

\begin{CodeChunk}
\begin{CodeInput}
R> plot(out, plot.type = "trend", trend.plot.args = list(
        ylab = "Luminescence (AU)",
        T.col = "blue", T.lwd = 2, main = ""),
        cex.lab = 1.3, col = "grey60")

R> lines(out.EP1$trend.est$T, col = "black", lwd = 2, lty = 2)
R> lines(out.EP4$trend.est$T, col = "red", lty = 3, lwd = 2)
R> lines(out.LA10$trend.est$T, col = "green", lty = 4, lwd = 2)

R> legend(10, 85000, c("EP10", "EP1", "EP4", "LA10"),
       lty = c(1, 2, 3, 4),lwd = c(2, 2, 2, 2),
       col = c("blue", "black", "red", "green"), bty = "n", cex = 1.2)
\end{CodeInput}
\end{CodeChunk}

\tcb{We compare the fits produced by the Daubechies Extremal Phase (EP) wavelets with 1, 4, and 10 vanishing moments, as well as the Daubechies Least Asymmetric (LA) Wavelet with 10 vanishing moments. The fits for the EP1 and EP4 wavelets appear to be noticeably worse than those for the EP10 and LA10, which show better adaptivity to the data and are almost identical. Heuristically, we can also compute the root mean squared error (RMSE) of the trend estimates, which identify the EP10 and LA10 wavelets as being indistinguishable (RMSEs of $3682.85$ and $3682.42$ respectively), and superior to the EP1 and EP4 wavelets (RMSEs of $4062.61$ and $3990.25$ respectively). The code for this is given below.}

\begin{CodeChunk}
\begin{CodeInput}
R> n <- length(celegensbio)
R> trends <- list(EP1 = out.EP1$trend.est$T,EP4 = out.EP4$trend.est$T,
               EP10 = out$trend.est$T, LA10 = out.LA10$trend.est$T)
R> trends.rmse <- lapply(trends,
                         function(y){sqrt(sum((y - celegensbio) ^ 2) / n)})
\end{CodeInput}
\end{CodeChunk}

\subsection{Accelerometer data}
\begin{figure}
    \begin{center}
        \includegraphics[width =0.9\textwidth]{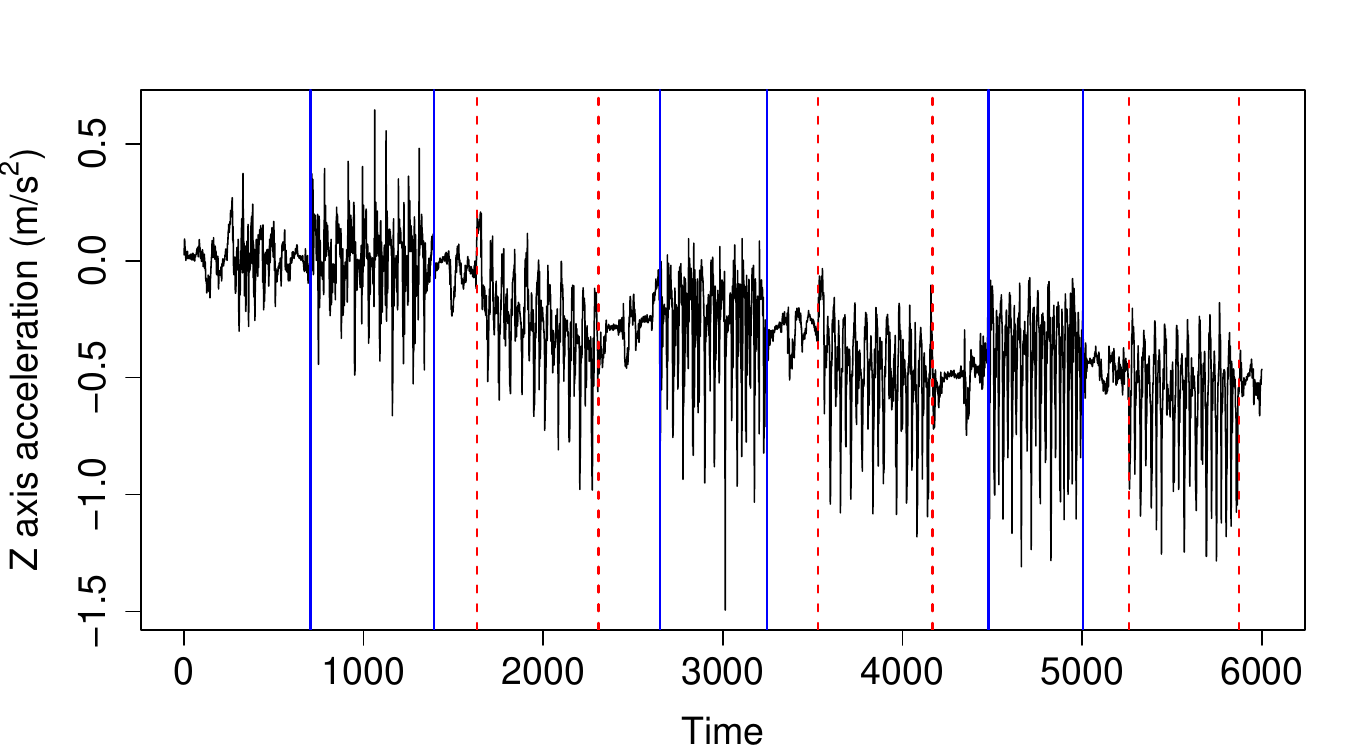}        
    \end{center}
    \caption{Z-axis acceleration time series. Data between blue solid vertical lines indicates when the participant was walking downstairs, data between red dashed vertical lines indicates walking upstairs.}\label{fig:accel}
\end{figure}

Accelerometers are used in a variety of fields including engineering, health, and marketing. Here we consider a dataset obtained from the UCI data repository \citep{UCIrepo} based on accelerometer readings from a smartphone \citep{Accel1} during an experiment on daily living activities.  Specifically, we analyse a section of the z-axis time series of length $n=6000$ from experiment 3 of user 2, which we denote \texttt{z.acc}. During this time period, the participant performed two activities several times; walking upstairs and walking downstairs. The data are shown in Figure~\ref{fig:accel}, where data between blue solid vertical lines indicate when the participant was walking downstairs, and data between red dashed vertical lines indicate walking upstairs. The data show a gradual downward trend, and periods of activity correspond to periods of greater variability. The accelerometer time series is available as \texttt{z.acc} within the \pkg{TrendLSW} package, and the associated activity labels (along with their start and end times) are available as \texttt{z.labels}.

We focus on investigating what insights the spectral estimate reveals about the activity of the participant. The following code performs TLSW estimation on \texttt{z.acc} and plots the result, where we have changed the default spectral smoothing to use the Epanechnikov kernel with a bin width of 150, corresponding to a time period of 3 seconds:

\begin{CodeChunk}
\begin{CodeInput}
R> data("z.acc")
R> data("z.labels")
R> z.acc.TLSW <- TLSW(z.acc, S.smooth.type = "epan", S.binwidth = 150)
R> plot(z.acc.TLSW, plot.type = "spec", 
        spec.plot.args = list(scaling = "by.level"))
\end{CodeInput}
\end{CodeChunk}

\begin{figure}
    \begin{center}
        \includegraphics[width =0.9\textwidth]{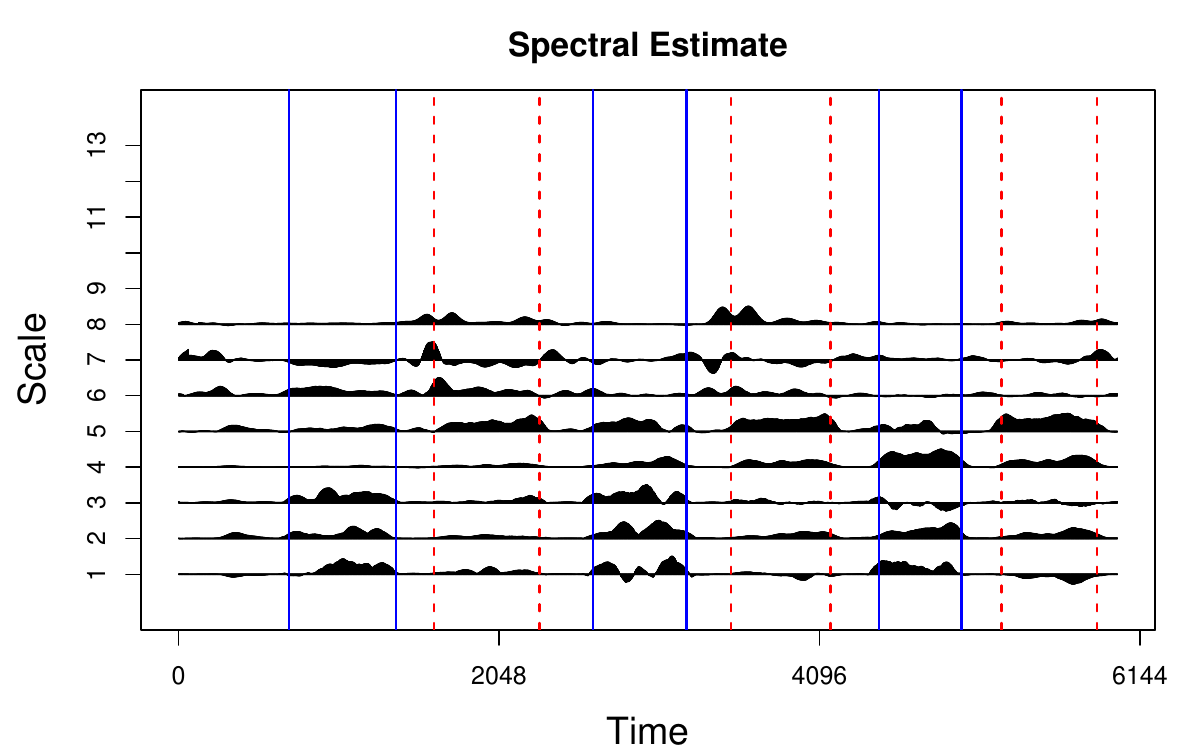}        
    \end{center}
    \caption{Spectral estimate for the accelerometer data from Figure~\ref{fig:accel}. Each level is scaled individually.}\label{fig:accel-spec}
\end{figure}

\begin{figure}
    \begin{center}
        \includegraphics[width =0.9\textwidth]{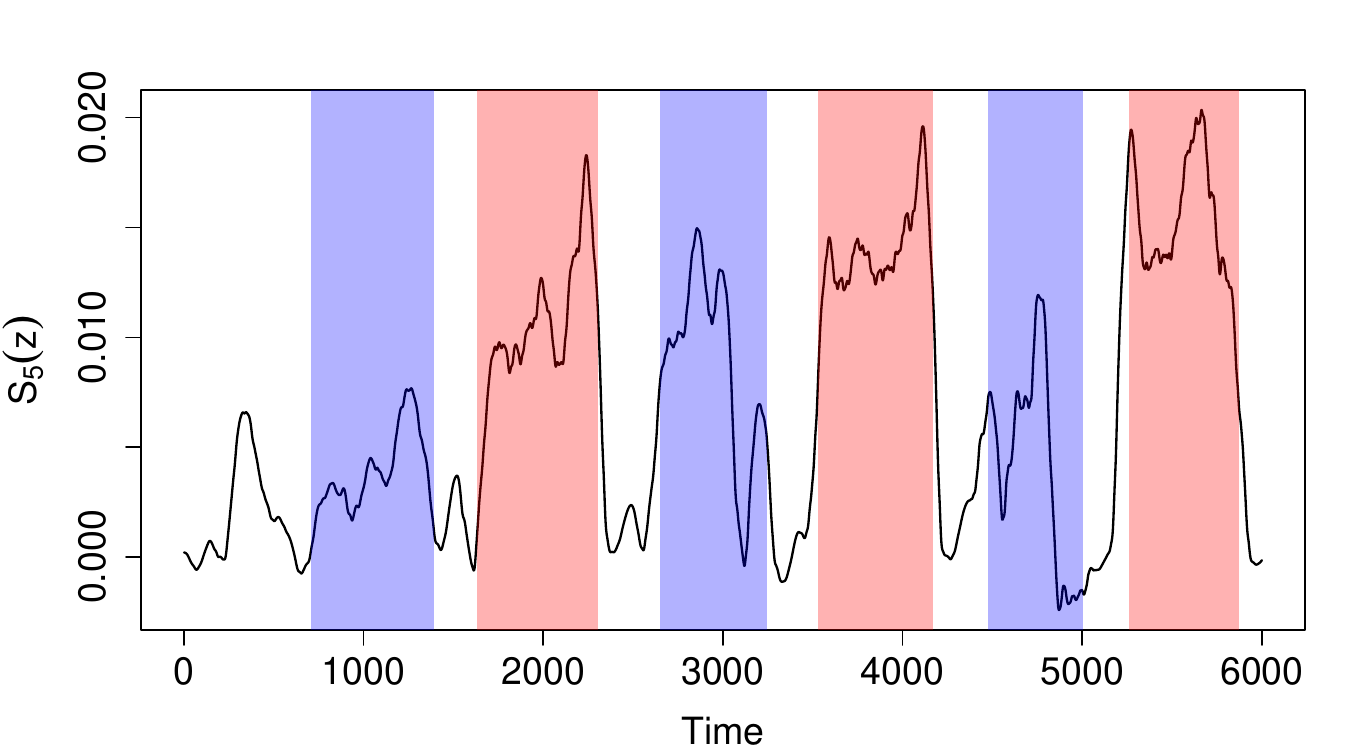}        
    \end{center}
    \caption{Spectral estimate at scale $j=5$ for the accelerometer data from Figure~\ref{fig:accel}. Periods of walking upstairs shown in blue, and periods of walking downstairs shown in red.}\label{fig:accel-spec5}
\end{figure}

Figure~\ref{fig:accel-spec} displays the spectral estimate of the time series, which has been scaled individually at each level. The spectral estimate closely correlates with the participant's activities. To highlight this, Figure~\ref{fig:accel-spec5} shows the spectral estimate at scale $j=5$ only, with periods of walking downstairs and walking upstairs superimposed onto the plot in blue and red respectively. We see that the periods of activity give rise to increased power in $S_5(z)$.

\section{Concluding remarks}\label{sec:conclusion}

Time series data with both time-varying mean and dependence structure are increasingly commonplace, and thus there is a growing need for implementations of statistically justified models and inference tools so that practitioners can derive principled insight from scientific data.  This article introduces the \pkg{TrendLSW} package in \proglang{R} to address this gap in currently available software, implementing recent methods for modelling nonstationary time series.  The functionality in the package allows users to simulate time series with first and second order nonstationarity, as well as estimate relevant quantities of interest, such as the trend and wavelet spectrum associated to time series.  

Whilst the \pkg{TrendLSW} package incorporates implementations of flexible estimation methods for the recent Trend-LSW model for users familiar with locally stationary wavelet processes, the package is designed with a broad end-user base in mind, and hence we provide an automatic way of analyzing nonstationary time series data, with few tuning parameters needed to be specified. Trend and dependence time series components are easily estimated and subsequently visualized, together with quantification of estimation uncertainty.  We have presented several case studies from different fields to demonstrate the package functionality, and hope the package will provide practical tools for data analysis in a wide range of scientific and industrial settings.

\tcb{Future development of \pkg{TrendLSW} will include adding further functionality including a predict method for \texttt{TLSW} objects.}

\section*{Acknowledgments}
Thanks to Alexandre Benedetto for collecting the data from the C. Elegans experiment and allowing us to provide it as open-source data within the package.

\bibliographystyle{apalike}
\bibliography{TrendLSW-bib}

\end{document}